\documentclass[conference]{IEEEtran}
\IEEEoverridecommandlockouts
\usepackage{cite}
\usepackage[switch]{lineno}

\usepackage{amsmath,amssymb,amsfonts}
\usepackage{algorithmic}
\usepackage{graphicx}
\usepackage{textcomp}
\usepackage{xcolor}
\usepackage[vlined,ruled,linesnumbered]{algorithm2e}
\usepackage{makecell}
\usepackage{subfigure}
\usepackage{mathtools}
\usepackage{pbox}
\usepackage{booktabs}
\usepackage{microtype}
\usepackage{flushend}

\usepackage{caption}
\def\BibTeX{{\rm B\kern-.05em{\sc i\kern-.025em b}\kern-.08em
    T\kern-.1667em\lower.7ex\hbox{E}\kern-.125emX}}
\begin{document}

\title{Faster CryptoNets: Leveraging Sparsity for Real-World Encrypted Inference\\
}


\author{
    \IEEEauthorblockN{Edward Chou\IEEEauthorrefmark{1}\thanks{authors \IEEEauthorrefmark{1} and \IEEEauthorrefmark{2} contributed equally}, Josh Beal\IEEEauthorrefmark{2}, Daniel Levy\IEEEauthorrefmark{3}, Serena Yeung\IEEEauthorrefmark{4},  Albert Haque \IEEEauthorrefmark{5}, Li Fei-Fei\IEEEauthorrefmark{7}}
    \IEEEauthorblockA{Department of Computer Science, Stanford University\\
    Email: 
\IEEEauthorrefmark{1}ejchou@cs.stanford.edu,
\IEEEauthorrefmark{2}joshbeal@cs.stanford.edu,
\IEEEauthorrefmark{3}danilevy@cs.stanford.edu,\\
\IEEEauthorrefmark{4}syyeung@stanford.edu,
\IEEEauthorrefmark{5}ahaque@cs.stanford.edu,
\IEEEauthorrefmark{6}feifeili@cs.stanford.edu}}

\maketitle

\begin{abstract}
Homomorphic encryption enables arbitrary computation over data while it remains encrypted.
This privacy-preserving feature is attractive for machine learning, but requires significant computational time due to the large overhead of the encryption scheme.
We present \textit{Faster CryptoNets}, a method for efficient encrypted inference using neural networks.
We develop a pruning and quantization approach that leverages sparse representations in the underlying cryptosystem to accelerate inference.
We derive an optimal approximation for popular activation functions that achieves maximally-sparse encodings and minimizes approximation error.
We also show how privacy-safe training techniques can be used to reduce the overhead of encrypted inference for real-world datasets by leveraging transfer learning and differential privacy.
Our experiments show that our method maintains competitive accuracy and achieves a significant speedup over previous methods.
This work increases the viability of deep learning systems that use homomorphic encryption to protect user privacy.
\end{abstract}


\section{Introduction}
As cloud-based machine learning services become more widespread, there is a strong need to ensure the confidentiality of sensitive healthcare records, financial data, and other information that enters third-party pipelines.
Traditional machine learning algorithms require access to raw data, which opens up potential security and privacy risks.
For some fields such as healthcare, regulations may preclude the use of external prediction services if the technology cannot provide the necessary privacy guarantees. 

In this work, we address the task of \textit{encrypted inference} for secure machine learning services.
We make the assumption that the third-party provider already has a trained model, as is common in ``machine learning as a service" paradigms.
Using cryptographic techniques, an organization such as a research hospital or fraud detection company will be able to offer prediction services to users while ensuring security guarantees for all parties involved.
We follow the procedure set by previous work \cite{gilad2016cryptonets, xie2014crypto} and employ homomorphic encryption (HE) to convert a trained machine learning model into a HE-enabled model.

\begin{figure}[t]
    \centering
    \includegraphics[width=1.0\linewidth]{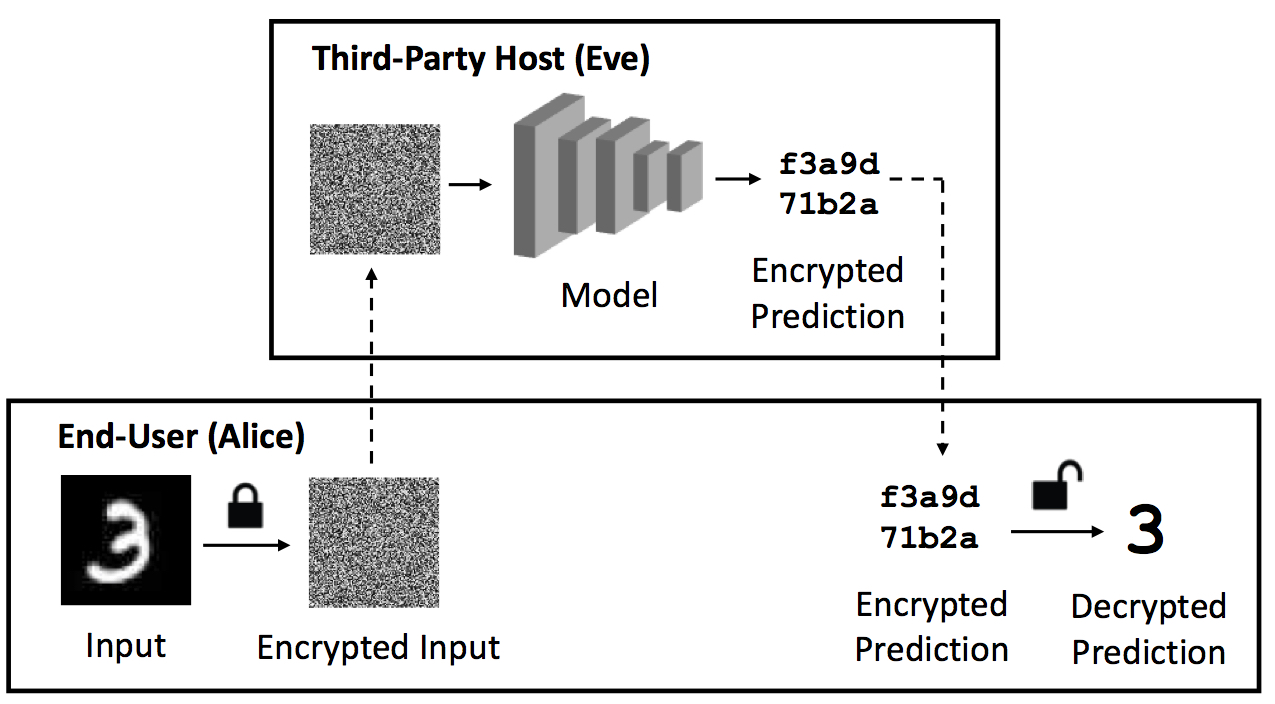}
    \vspace{-5mm}
    \caption{\textbf{Encrypted machine learning as a service paradigm.} Dashed lines indicate data transfer. The end-user (Alice) encrypts her sensitive data and sends it to a third-party host (Eve). Since Alice owns the private key, Eve cannot decrypt the input nor output prediction. Eve produces an encrypted prediction which is returned to Alice. Privacy is preserved in the entire pipeline for both inputs and outputs.}
    \label{fig:pull}
\end{figure}

Homomorphic encryption \cite{rivest1978data} allows a machine learning model to perform calculations over encrypted data. 
By design, the output prediction is also encrypted, which prevents the input or output from leaking information to the model's host.
As show in Figure \ref{fig:pull}, the model does not decrypt the data nor is the private key needed 
\cite{brakerski2014efficient}.



Several challenges prevent widespread adoption of encrypted machine learning.
A major bottleneck is computational complexity.
Inference on plain networks is performed in the orders of milliseconds,
while encrypted networks require minutes or hours per example \cite{gilad2016cryptonets, hesamifard2017cryptodl}.
Also, the reduced arithmetic set of HE prevents the use of modern activation functions \cite{chabanne2017privacy}.
necessitating the use of simpler lower-performance functions.

\subsection{Contributions}

We propose \textit{Faster CryptoNets} -- a method for encrypted inference on the order of seconds.
This is a significant improvement over existing state-of-the-art, which performs inference on the order of minutes.
Our contributions accelerate the homomorphic evaluation of deep learning models on encrypted data using sparse representations throughout the neural network.
Additionally, we are able to efficiently approximate modern activation functions.  
Finally, we show how this technique can be combined with private training techniques in a plausible real-world scenario.

By intelligently pruning the network parameters, we can avoid many multiplication operations -- a major contributor to computational complexity. 
We can progressively quantize the remaining network parameters such that the plaintext encodings achieve maximum sparsity.
Also, given that the activation function is the single most expensive operation of the network, we derive an optimal, quantized polynomial approximation to the activation function also with maximally-sparse encodings.
We empirically show a significant improvement in the runtime of the network on MNIST.
We perform additional experiments on larger datasets to demonstrate the viability and performance gain on practical tasks.
We use a feature-extraction based framework to reduce the number of layers requiring encrypted computation, while using differentially private training to achieve competitive accuracy on real-world datasets.

\section{Related Work}\label{sec:related_work}
\subsection{Related Work}

Privacy-preserving machine learning models attempt to address computation and statistical modeling of private data \cite{agrawal2000privacy}.
Privacy is preserved when two conditions are met: (i) the end-user learns nothing about the model and (ii) the model learns nothing from the data \cite{brickell2009privacy}.
Differential privacy, multi-party computation (MPC), and homomorphic encryption are different methods to preserve privacy.

Differential privacy allows statistics to be computed over a dataset without revealing information about individual records \cite{dwork2008differential, chaudhuri2011differentially}.
A common method is to apply noise to individual examples to obfuscate statistical differences that might be distinguishable \cite{papernot2016semi}.
However, differential privacy is better suited for the training phase. During test-time, adding noise to a single example may change the prediction.

Secure multi-party computation enables multiple parties to jointly compute a function over their inputs while keeping their inputs private.
This has been explored using Garbled Circuits \cite{yao1986generate} in the works of \cite{rouhani2017deepsecure, kolesnikov2008improved} and \cite{mohassel2017secureml}.
These methods often involve a high communication complexity with significant bandwidth costs. 

Fully homomorphic encryption (FHE) was proposed by \cite{gentry2009fully} and allows anyone to compute over encrypted data without decrypting it \cite{naehrig2011can}.
A weaker version of FHE, termed leveled homomorphic encryption (LHE) permits a subset of arithmetic operations on a depth-bounded arithmetic circuit \cite{brakerski2014efficient}.
While HE has been explored for machine learning applications, many works focus on simpler models such as linear \cite{hall2011secure}, logistic \cite{cheon2017privacy} and ridge regression \cite{irene2017ridge}.
CryptoNets \cite{gilad2016cryptonets} was one of the first works to implement HE in a neural network setting.
More recently, \cite{chabanne2017privacy} and \cite{hesamifard2017cryptodl} extended this to deeper network architectures and developed additional polynomial approximations to the activation function that leveraged batch normalization for stability.

Other works have explored the broader use of polynomial activation functions.
\cite{piazza1992artificial} and \cite{ma2005constructive} used a polynomial function in the non-encrypted domain to some success.
The original theory dates back to \cite{hornik1991approximation} who argues that as long as the activation function is arbitrarily bounded and non-constant, the neural network is a universal approximator.
Some prior work even suggests that neural networks equipped with polynomial functions have the same representational power as their non-polynomial counterparts \cite{gautier2016globally, livni2014computational}.
In \S\ref{sec:prelim} and \S\ref{sec:method}, we explore these ideas in greater detail.

Recent works have proposed techniques that accelerate neural network inference on encrypted data.
Sanyal et. al. \cite{sanyal2018tapas} use sparsification techniques via binarized neural networks which achieves a similar speedup of around 30x wall-clock time as our technique on MNIST.
Florian et. al. \cite{florian2017fastdiscretize} opt for an approach that leverages scale invariance to allow unrestricted depth of neural networks.  The technique we propose is distinct from these approaches due to its use of the encoding scheme to accelerate multiplicative operations, in contrast to the previous approaches which bypass expensive operations using the sign activation function. Our approach is advantageous in that it is more compatible with common neural network components; sign activation functions are known to cause difficulty with convergence, and the scale invariant \cite{florian2017fastdiscretize} precludes the use of convolutional layers.  We do not present detailed comparisons to these works in our analysis due to these fundamental architectural differences, and opt for a direct comparison to CryptoNets to clearly demonstrate in which layers and with which operations are our speedups derived from.

\section{Threat Model}\label{sec:threat_model}
Machine Learning as a Service (MLaaS) \cite{bae2018securityprivacyissues} is a framework where cloud providers offer machine learning training of inference hosted on the cloud.  In our scenario we will be considering a MLaaS inference pipeline, where users send data to a remote server and receive predictions performed by machine learning models.  The machine learning model is pre-trained on a proprietary dataset.

A universal threat in multi-party situations is the inherent risk of data transmission, either by interception or side-channel attacks.  This threat can be mitigated to a large extent by using strong cryptographic and signature protocols to protect the data in-transmission. However, a concern that is much harder to alleviate involves the threat of the cloud host collecting and utilizing the transmitted data without authorization \cite{bae2018securityprivacyissues}.  In a naive scheme, a user sends encrypted data to the cloud, but also has to provide a key to the server to decrypt the data and compute a output with a machine learning algorithm before sending the encrypted prediction back to the user.  The cloud host must have access to the plain data, and it is hard to guarantee or prove to the user that the data is not kept on the server, where it can either be sold to third-parties or be stolen by attackers who gain access to the data.

Homomorphic encryption provides a solution to both problems.  By design, the transmitted data is protected using a strong encryption scheme.  It also enables "oblivious inference", where a cloud host operates on data that it is oblivious to.  If the service provider is only allowed to compute on the encrypted data to compute an encrypted output without ever decrypting the data at any step, it will never have access to the plain data, guaranteeing data privacy from the cloud provider.

\section{Preliminaries}\label{sec:prelim}

A \textit{homomorphism} is a structure-preserving transformation between two algebraic structures, which can be leveraged by cryptosystems to allow for arithmetic operations on encrypted data. 
Let $G_q$ be a cyclic group of order $q$ with generator $g$. Let $h \in G_q$ be randomly sampled as the public key.
Consider the ElGamal encryption scheme \cite{elgamal1985public}, which uses a map $\Phi: G_q \rightarrow  G_q \times G_q$ such that $\Phi(m) := (g^r, mh^r)$ for random $r$. 
The map $\Phi$ preserves the multiplicative structure of the integers such that $\Phi(m_1 \times m_2) = \Phi(m_1) \otimes \Phi(m_2) = (g^{(r_1+r_2)}, (m_1 \times m_2)h^{(r_1+r_2)}$) where $\otimes$ is the multiplication operation in $G_q \times  G_q$.

The leveled homomorphic encryption scheme that we present below has a more complex algebraic structure, and supports both additive and multiplicative homomorphisms, but this example can serve as a basis for understanding the role of homomorphic encryption in our network design.

\subsection{Notation}
Let $R_k$ denote the polynomial ring $\mathbb{Z}_k[x]/(x^n + 1)$.
We let $x \leftarrow S$ denote uniformly random sampling of $x$ from an arbitrary set $S$, and $\lfloor \frac{t}{q} p \rceil$ denote a coefficient-wise division and rounding of the polynomial $p$ with respect to integer moduli $t$ and $q$.
Let $[p]_q$ denote the reduction of the coefficients of the polynomial $p$ modulo $q$, and let $\Delta$ denote $\lfloor {q/t} \rfloor$.

\subsection{Encryption Scheme.}
Bajard et al. \cite{bajard2016full} proposed an encryption scheme, FV-RNS, which is a residue number system (RNS) variant of the FV encryption scheme. 
In FV-RNS, plaintexts are elements of the polynomial ring $R_t$, where $t$ is the plaintext modulus and $n$ is the maximum degree of the polynomial, which is commonly selected to be one of $\{1024, 2048, 4096, 8192, 16384, 32768\}$. 
The plaintext elements are mapped to multiple ciphertexts in $R_q$ in the encryption scheme, with $q \gg t$ as the ciphertext coefficient modulus.
For any logarithm base $\beta$, let $\ell = \lfloor {\log_\beta q} \rfloor$ be the number of terms in the base-$\beta$ decomposition of polynomials in $R_q$ that is used for relinearization.

Let $\chi$ denote the truncated discrete Gaussian distribution.
The secret key is generated as $s \leftarrow R_3$ with coefficients $s_i \in \{ 0, 1, -1\}$.
The public key $(p_0, p_1)$ is generated by sampling $p_0 \leftarrow R_q$ and $e' \leftarrow \chi$ and constructing $p_1 = [−(s p_0 + e')]_q$.
The evaluation keys $(a_i, g_i)$ are generated by sampling $a_i \leftarrow R_q$ and constructing $g_i=[−(a_i s + ie') + \beta^i s^2]_q$ for each $i \in \{0, ..., \ell\}$.

A plaintext $m \in R_t$ is encrypted by sampling $u \leftarrow R_3$ with coefficients $u_i \in \{ 0, 1, -1\}$ and $e_1,e_2 \leftarrow \chi$, and letting $(c_0,c_1)=([\lfloor {q/t} \rfloor m+p_0u+e_1]_q,[p_1u+e_1]_q)$.
A ciphertext $(c_0,c_1) \in R_q \times R_q$ is decrypted as $m=[\lfloor \frac{t}{q} [c_0+c_1s]_q \rceil]_t \in R_t$.

\subsection{Arithmetic}.
The addition of two ciphertexts $(c_0,c_1)$ and $(d_0,d_1)$ is $(c_0+d_0,c_1+d_1)$.
The multiplication of two ciphertexts $(c_0,c_1)$ and $(d_0,d_1)$ occurs by constructing
\begin{equation*}
    c_0' = \left[\left\lfloor \frac{t}{q} [c_0 d_0] \right\rceil\right]_q \textrm{ ,\quad } c_1' = \left[\left\lfloor \frac{t}{q} [c_0 d_1 + c_1 d_0] \right\rceil\right]_q,
\end{equation*}
\begin{equation*}
   \textrm{ and \,} c_2' = \left[\left\lfloor \frac{t}{q} [c_1 d_1] \right\rceil\right]_q .
\end{equation*}
We express $c_2'$ in base $\beta$ as $c_2' = \sum_{i=0}^{\ell} c_2'^{(i)} \beta^i$. We then let $r_0=c_0'+\sum_{i=0}^{\ell} a_i c_2'^{(i)}$ and $r_1= c_1'+\sum_{i=0}^{\ell} g_i c_2'^{(i)}$, which forms the product ciphertext $(r_0, r_1) \in R_q \times R_q$.

The addition of ciphertext $(c_0,c_1)$ and plaintext $m$ is the ciphertext $(c_0+\Delta m,c_1)$. The multiplication of ciphertext $(c_0,c_1)$ and plaintext $m$ is the ciphertext $(mc_0,mc_1)$.

The advantage of the residue number system variant is that the coefficient modulus $q$ can be decomposed into several small moduli $q_1, ..., q_k$ to avoid multiple-precision operations on the polynomial coefficients in the homomorphic operations, which improves the efficiency of evaluation.

\subsection{Integer Encoder}.
To encode real numbers involved in the computation, we choose a fixed precision for the values (15 bits) and scale each value by the corresponding power of 2 to get an integer for use with the encoder described below. After decryption, we can divide by the accumulated scaling factor to obtain a real value for the prediction.
The encoder consists of a base-2 integer encoder \cite{seal23}.
For a given integer $z$, consider the binary expansion of $|z| = z_{n-1}...z_1 z_0$.
The the coefficients $b_i$ of the polynomial $f(x) = \sum_{i=0}^{n - 1} b_ix^i$ in the plaintext ring are $z_i$ if $z_i \geq 0$ otherwise $b_i=t-z_i$.


\section{Method}\label{sec:method}


\subsection{Sparse Polynomial Multiplication}

The convolutional and fully connected layers of a neural network require a substantial number of multiplications involving both the ciphertext inputs and the plaintext parameters of the model.
Each operation involves computing the product of two polynomials with up to $n$ nonzero coefficients.
While a brute-force implementation would require $\mathcal{O}(n^2)$ time to complete, homomorphic encryption methods are able to accomplish this in $\mathcal{O}(n\log{}n)$ when certain conditions are met.
Assuming that the coefficient modulus $q$ is chosen such that $(q-1)$ is divisible by $2n$, we can invoke the Number Theoretic Transform to achieve $\mathcal{O}(n\log{}n)$ \cite{harvey2014faster}.

Our contributions leverage the following insight: a substantial improvement in efficiency occurs when the plaintext multiplier  $z = \pm 2^k$ for some $k \in \mathbb{Z}$.
The polynomial that encodes this integer is $b_k x^k$, a monomial multiplier.
For such parameters, sparse polynomial multiplication \cite{akleylek2016efficient} has been shown to use $\mathcal{O}(n)$ coefficient multiplications and modular reductions (see Algorithm \ref{alg:mult}).

\begin{algorithm}[!htbp]
   \caption{Sparse Plaintext-Ciphertext Multiplication}
   \label{alg:mult}
\begin{algorithmic}
   \STATE {\bfseries Input:} ciphertext $c = \sum_{i=0}^{n - 1} c_ix^i$
   plaintext $b=b_k x^k$
   \FOR{$i=0$ {\bfseries to} $n-1$}
   \STATE Initialize $j = i + k$.
   \IF{$j \geq n-1$} 
   \STATE $d_{j-(n-1)} = c_i (t - b_k)$
   \ELSE
   \STATE $d_j = c_i b_k$
   \ENDIF
   \ENDFOR
   \STATE {\bfseries Output:} ciphertext $d = \sum_{i=0}^{n - 1} d_ix^i$
\end{algorithmic}
\end{algorithm}

\subsection{Network Pruning and Quantization}\label{sec:pruning}

The parameters of a neural network can be iteratively removed and clustered without affecting accuracy.
\cite{han2015deep} developed a compression method that leverages these techniques.
Since then, new pruning and quantization techniques have been proposed \cite{luo2017thinet}. 
We leverage these techniques to reduce the number of weights that contribute to the multiplication count, and convert the weights to powers of 2, which have sparse polynomial representations that reduce the cost of each multiplication. Together, these lead to significant reductions in inference time.

We first train a pruned version of the network with Dynamic Network Surgery (DNS) \cite{guo2016dynamic} that incorporates connection splicing.
The remaining network parameters are quantized to powers of 2 following the incremental network quantization (INQ) procedure proposed by \cite{zhou2017incremental}.
The INQ method consists of an iterative quantization strategy to preserve the original inference accuracy.

For each layer $\ell$, the layer's parameters $\mathbf{W_\ell}$ have a corresponding binary pruning mask $\mathbf{T_\ell}$.
The elements of the binary pruning mask $\mathbf{T_\ell}$ get updated during gradient descent according to a discriminative measure of parameter importance $h_\ell$, typically incorporating a magnitude-based measure $s$ such as $s = \max|\mathbf{W_\ell}|$.
We define $n_1$ and $n_2$, which will help bound our quantized values for each layer:
\begin{equation*}
    n_1 = \left\lfloor \log_2 \frac{4s}{3} \right\rfloor \textrm{ and \,} n_2 = n_1 + 1 - 2^{\frac{k-1}{2}},
\end{equation*} 
where $k$ is used to restrict the set of powers for our desired bitwidth. We use $k=5$.
Note that $n_1 \geq n_2$ and
$$P = \{-2^{n_1}, ... , -2^{n_2}, 0, 2^{n_2}, ..., 2^{n_1} \}$$
is the set of possible quantized values for the parameters of layer $\mathbf{W_\ell}$ in the network. 
We define a monotonically increasing weight partition schedule using the discriminative measure $h_\ell$ to progressively quantize the weights.
For example, one can quantize 50\% of the weights, then 75\%, then 87.5\%, then 100\%, retraining the other non-quantized weights at each step of the quantization procedure.


\subsection{Approximating the Activation Function}\label{sec:approx_activation}

Using our pruning and quantization scheme from \S\ref{sec:pruning}, our next contribution lies in finding the optimal polynomial approximation for any activation function given the constraint that the coefficients must be a power of 2.
The activation function of a neural network is critical for convergence \cite{glorot2011deep} and has been thoroughly explored in literature \cite{ramachandran2017swish}.
With the goal of encrypted network inference, we must find an approximation which balances approximation error with practical usability.
Inspired by \cite{brisebarre2006computing}, we find the best polynomial approximation.

\textbf{Polynomials}. Let $x \in \mathbb{R}$ and let $f: \mathbb{R} \rightarrow \mathbb{R}$ denote the activation function.
Our task is to approximate $f$ with a polynomial $p^*: \mathbb{R} \rightarrow \mathbb{R}$ where $p^*(x) = p^*_0 + p_1^* x + \cdots + p^*_n x^n$ subject to the constraint that each coefficient is a power of 2.
Define $\mathcal{P}^{(2)}_n$ as the set of all polynomials of degree less than or equal to $n$, such that all coefficients are base-2.
That is, $\mathcal{P}^{(2)}_n = \{  2^{a_0} + 2^{a_1} x + \cdots + 2^{a_n} x^n , a_i \in \mathbb{Z} \}$.
Let $p$ be the minimax approximation to $f$ on some interval $[-a, a]$.
Let $\hat{p}$ be the same as $p$, but with all coefficients rounded to the nearest $2^k$ where $k \in \mathbb{Z}$.
Note, $\hat{p} \in \mathcal{P}^{(2)}_n$.

\textbf{Maximum Error \& Minimax}. The maximum difference (i.e., error) $\delta$ between two functions $g$ and $h$ is $\delta(g, h) = \max_{x\in [-a,a]} | g(x) - h(x) |$.
This provides a strong bound on the optimal polynomial approximation error $\delta(f, p^*)$ where $\delta(f, p) \leq \delta(f, p^*) \leq \delta(f, \hat{p}).$
We state minimax problem as follows. For a given activation function $f$, we seek to find the best polynomial $p^* \in \mathcal{P}^{(2)}_n$ such that,
\begin{equation}
    \delta(f, p^*) = \min_{q \in \mathcal{P}^{(2)}_n} \delta(f, q)
\end{equation}
subject to the constraint,
\begin{equation}\label{eq:constraint}
    \delta(f, p^*) \leq K \textrm{ where } K \geq \delta(f, p).
\end{equation}

\textbf{Finite Number of Solutions.}
Let $d, n \in \mathbb{N}$, and $D = \{0,...,d\}$. For $\forall i \in D$, let $x_i, l_i, u_i \in \mathbb{R}$ such that $x_j \neq x_k$ if $j\neq k$. We can construct a bounded polyhedron,
\begin{equation*}
    \mathcal{B} = \left\{ (\alpha_0, ..., \alpha_n) \in \mathbb{R}^{n+1} \; \Bigg\vert \; l_i \leq \sum\limits_{j=0}^{n} \alpha_j x^j_i \leq u_i, \forall i \in D  \right\}
\end{equation*}
where each $(\alpha_0,...,\alpha_n)$ tuple represents any polynomial $q \in \mathcal{P}^{(2)}_n$, and where $\alpha_i$ represents the degree $i$ coefficient.
\cite{brisebarre2006computing} show that the number of polynomials satisfying Equation \ref{eq:constraint} is finite if the polynomials are contained in $\mathcal{B}$.
They also proposed an efficient scanning method to find the optimal polynomial approximation $p^*$.
Equipped with our new-found approximation $p^*$, we can evaluate the effectiveness of $p^*$ as an activation function in both non-encrypted and encrypted domains.

\section{Experiments}


\subsection{Wall-clock Runtime}
The runtime refers to the wall-clock time required to perform inference on an encrypted image.  This metric is the default metric reported in previous encrypted works.  However, the wall-clock time is an imperfect metric for measuring improvements in encrypted inference. It is hardware-dependent, varying greatly depending on the available memory and computational power of the device, and it is also possibly encryption-scheme dependent, with even the same encryption algorithm being implemented differently across libraries.  In the next sections, we introduce a metric to evaluate our methods using hardware-independent metrics.

\subsection{Explanation of HOPs}\label{sec:exp_setting}



We report the number of homomorphic operations (HOPs) of our inference network.
This is in contrast to previous work \cite{gilad2016cryptonets,chabanne2017privacy,hesamifard2017cryptodl}, which measured either throughput or wall-clock time -- both of which are highly dependent on hardware specifications and software parallelization, and are not entirely reliable measures.
The HOPs metric is similar to the FLOPs (floating point) metric used in scientific computing.

A homomorphic operation is defined as addition or multiplication involving a ciphertext, a plaintext, or both.
The four classes of HOPs are (i) plaintext-ciphertext addition, (ii) ciphertext-ciphertext addition, (iii) plaintext-ciphertext multiplication, and (iv) ciphertext-ciphertext multiplication.
While the exact implementation of HOPs may vary, we believe HOPs are a hardware-independent metric for performance analysis that enable a better comparison of models for encrypted inference, demonstrating whether speedup occurs due to decreased number of operations or due to algorithmic speedup.

It is important to note that the different HOPs classes vary in cost.  In general, multiplicative operations are significantly more expensive than additive operations, with ciphertext-ciphertext multiplications being the most costly operations found in neural networks.  Throughout our analysis, we break down our HOPs into separate operations, following the rule-of-thumb that reducing multiplicative HOPs outweighs the cost of adding additive HOPs.

\subsection{Datasets}
We use the MNIST dataset of handwritten digits \cite{lecun1995convolutional} which contains $28 \times 28$ grayscale images of Arabic numerals 0 to 9 (i.e., 10-class classification task), which has a standard split of 50,000 training images and 10,000 images test set images.
While MNIST is arguably a simple dataset, it has remained the standard benchmark for homomorphic inference tasks \cite{gilad2016cryptonets, hesamifard2017cryptodl}.

\subsection{Network Architecture}
The network architecture used for MNIST inference is presented below.
The architecture itself is a slight variant of the CryptoNet \cite{gilad2016cryptonets} architecture that incorporates batch normalization layers to support a greater variety of activation functions.  The multiplicative depth is unchanged.
As shown in Figure \ref{fig:approx}, our approximation error is minimized close to zero.
Batch normalization encourages the pre-activation values to fit in this range.
As confirmed in \cite{chabanne2017privacy} and \cite{hesamifard2017cryptodl},  by reducing the variance in the input values to the activation layer, the approximation error of the network decreases.
Overall, our model is a convolutional neural network \cite{lecun1995convolutional} consisting of convolutional layers, activation functions, scaled average pooling, batch normalization, and fully-connected layers.

1. {\em Convolutional Layer.} The input image is 28 x 28. There are 20 kernels of size 5 x 5, with stride 2, and padding of 1.

2. {\em Batch Normalization Layer.} This layer applies the batch normalization weights and biases to each input value.

3. {\em Activation Layer.} This layer applies the approximate activation function to each input value.

4. {\em Scaled Average Pool Layer.} This layer has 3 x 3 windows, with a stride of 2, padding of 1, and output size of 5 x 13 x 13.

5. {\em Convolutional Layer.} This layer has 50 kernels of size 20 x 5 x 5, with a stride of 1, and zero padding.

6. {\em Scaled Average Pool Layer.} This layer has 3 x 3 windows and a stride of 2, padding of 1, and output size of 50 x 5 x 5.

7. {\em Fully-Connected Layer.} This layer has parameters of size 1250 x 100 for matrix multiplication with respect to inputs.

8. {\em Batch Normalization Layer.} This layer applies the batch normalization weights and biases to each input value.

9. {\em Activation Layer.} This layer applies the approximate activation function to each input value.

10. {\em Fully-Connected Layer.} This layer has parameters of size 100 x 10 for matrix multiplication with respect to inputs.

\subsection{Encryption Scheme}.
The parameters for the FV-RNS encryption scheme are: coefficient count of $n = 8192$, plaintext moduli of $t_1$ = 1099511922689 and $t_2$ = 1099512004609. The values of $q_i$ are selected for 128-bit security ($\log q$ = 219). This choice of coefficient modulus $q$ meets the security standards established by the Homomorphic Encryption Standardization Workshop \cite{albrecht2018homomorphic}.

\subsection{Hardware/Software Setup}.
The machine used for the MNIST experiments has an Intel Core i7-5930K CPU at 3.5 GHz with 48 GB RAM on Ubuntu 17.10.  The HE library was SEAL v2.3.0-4 \cite{seal23}, modified by us to support our proposed method.

\subsection{Optimization Hyperparmeters}
We provide the hyperparameter settings used to train our non-encrypted network.
A batch size of 64 was used and the model was trained for 30 epochs.
The learning rate schedule was initialized at $\lambda=0.008$ with a step size of 10 epochs and $\gamma=0.1$.
The model was trained with stochastic gradient descent with a momentum of 0.9.
For the square function, gradients were clipped at 0.25. He weight initialization was used for the convolutional layers.

\subsection{Dynamic Network Surgery Hyperparemeters}
We report the hyperparameters of our dynamic network surgery operations.  The sparsity denotes the final fraction of non-pruned connections over the total connections c-rate denotes compression rate used to set the threshold of importance before removing a connection.

We report metrics for each layer. The conv1 layer had a sparsity of 0.1440 and c-rate of 1.5. The conv2 layer had a sparsity of 0.0701 and c-rate of 1.65. The dense-fc1 layer had a sparsity of 0.0568 and c-rate of 1.65. The dense-fc2 layer had a sparsity of 0.1480 and c-rate of 1.5. All layers stopped at iteration 10,000.

\subsection{Approximation Results}\label{sec:approx}

\begin{figure*}[!htbp]
    \centering
    \subfigure[Swish$^1$: $f(x) = x / (1 + e^{-x})$]{
        \includegraphics[width=1.7in]{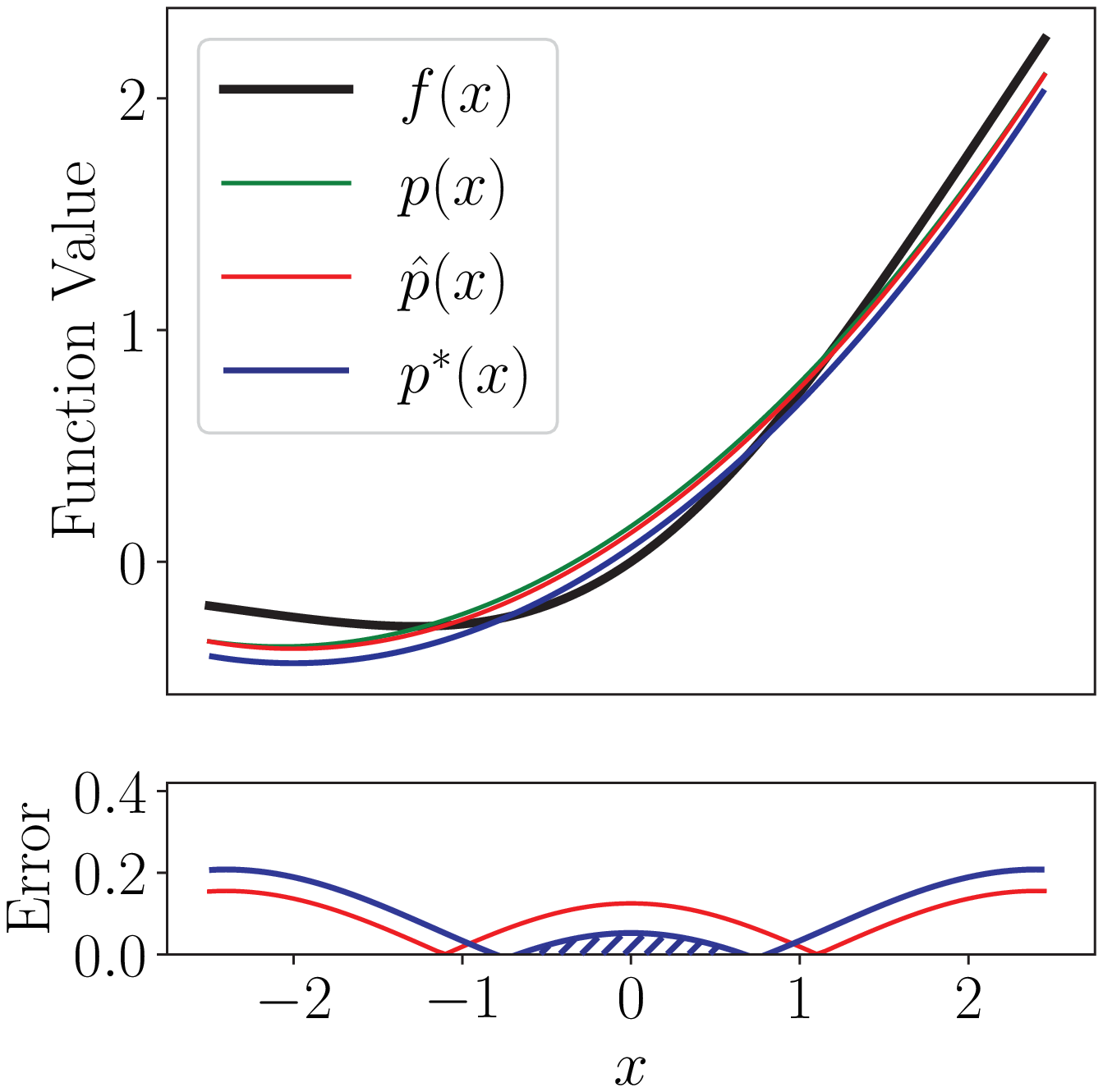}
        \label{fig:approx_swish}
    }
    \subfigure[ReLU: $f(x) = \max \{0, x\}$]{
        \includegraphics[width=1.7in]{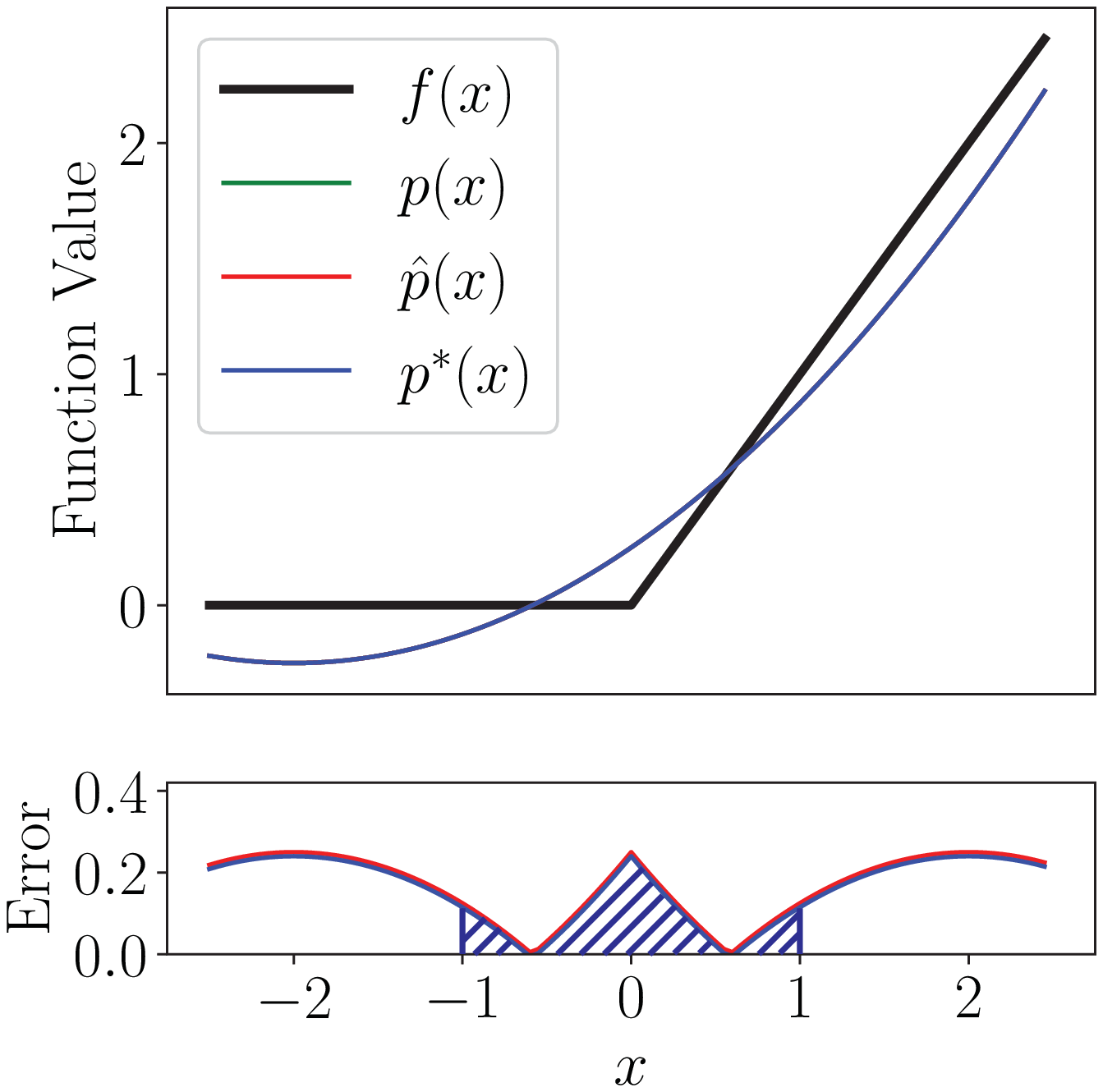}
        \label{fig:approx_relu}
    }
    \subfigure[Softplus: $f(x)=\log(1 + e^x)$]{
        \includegraphics[width=1.7in]{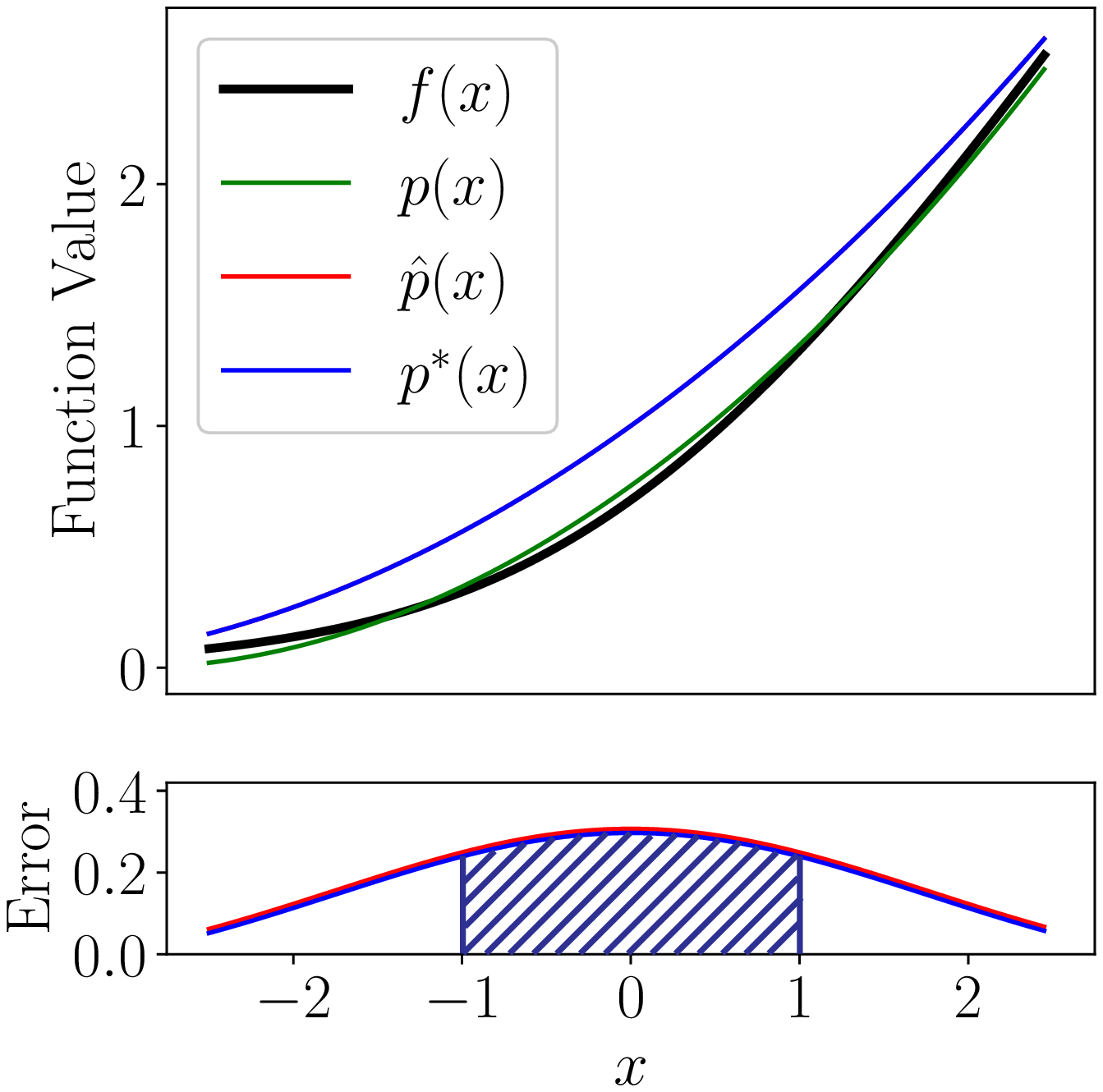}
        \label{fig:approx_softplus}
    }
    \caption{Approximation results (non-encrypted). (Top) Different approximation methods. The original activation function $f(x)$ is plotted with three approximations: the minimax estimate $p(x)$, the rounded minimax estimate $\hat{p}(x)$, and the our method -- the quantized minimax approximation $p^*(x)$. (Bottom) Error of our method. Our method $p^*(x)$ is compared to the baseline $\hat{p}(x)$. The blue shaded area corresponds to the post-batch normalization region during the training procedure.}
    \label{fig:approx}
\end{figure*}

Prior work suggests that neural networks equipped with polynomial functions have the same representational power as their non-polynomial counterparts \cite{gautier2016globally,livni2014computational}.
\textit{Faster CryptoNets} uses quadratic activation functions that approximate modern activations with varying degrees of complexity and expressivity. 
Our proposed method allows us to construct an optimal, quantized polynomial approximation of any arbitrary function.
In our experiments, we consider ReLU \cite{glorot2011deep}, Softplus \cite{dugas2001incorporating}, and Swish \cite{ramachandran2017swish}.
We model all activation functions with a 2\textsuperscript{nd} degree polynomial.
While higher-degree polynomials can decrease the approximation error, higher-degree polynomials also require more HOPs.

We note that \cite{ghodsi2017safetynets} showed how the gradient of the square function can be large.
Their solution was to apply gradient clipping to improve model convergence.
While this is a viable solution for recurrent networks \cite{chung2014empirical}, clipping gradients in a shallow network (such as ours) may indicate model instability and may not work for deeper variants.
To avoid this, we do not use the square activation function.

\subsection{Polynomial Approximation Equations}
We list the polynomial approximations to the Swish, Softplus, and ReLU activation functions.\\

Swish 
\begin{itemize}
    \item Minimax: $p = 0.12050344x^2 +  0.5x + 0.153613744$
    \item Rounded Minimax: $\hat{p} = 2^{-3}x^2 + 2^{-1}x + 2^{-3}$
    \item Quantized: $p^* = 2^{-3}x^2 + 2^{-1}x + 2^{-4}$
\end{itemize}

Softplus 
\begin{itemize}
    \item Minimax: $p = 0.082812671x^2 +  0.5x + 0.75248$
    \item Rounded Minimax: $\hat{p} = 2^{-4}x^2 + 2^{-1}x +  2^{0}$
    \item Quantized: $p^* = 2^{-4}x^2 + 2^{-1}x +  2^{0}$
\end{itemize}

ReLU 
\begin{itemize}
    \item Minimax: $p = 0.125x^2 +  0.5x + 0.25$
    \item Rounded Minimax: $\hat{p} = 2^{-3}x^2 + 2^{-1}x +  2^{-2}$
    \item Quantized: $p^* = 2^{-3}x^2 + 2^{-1}x +  2^{-2}$
\end{itemize}

\subsection{Error Minimization}

\begin{figure*}[t]
    \centering
    \hspace{-3mm}
    \subfigure[Convolution Layer]{
        \includegraphics[width=1.4in]{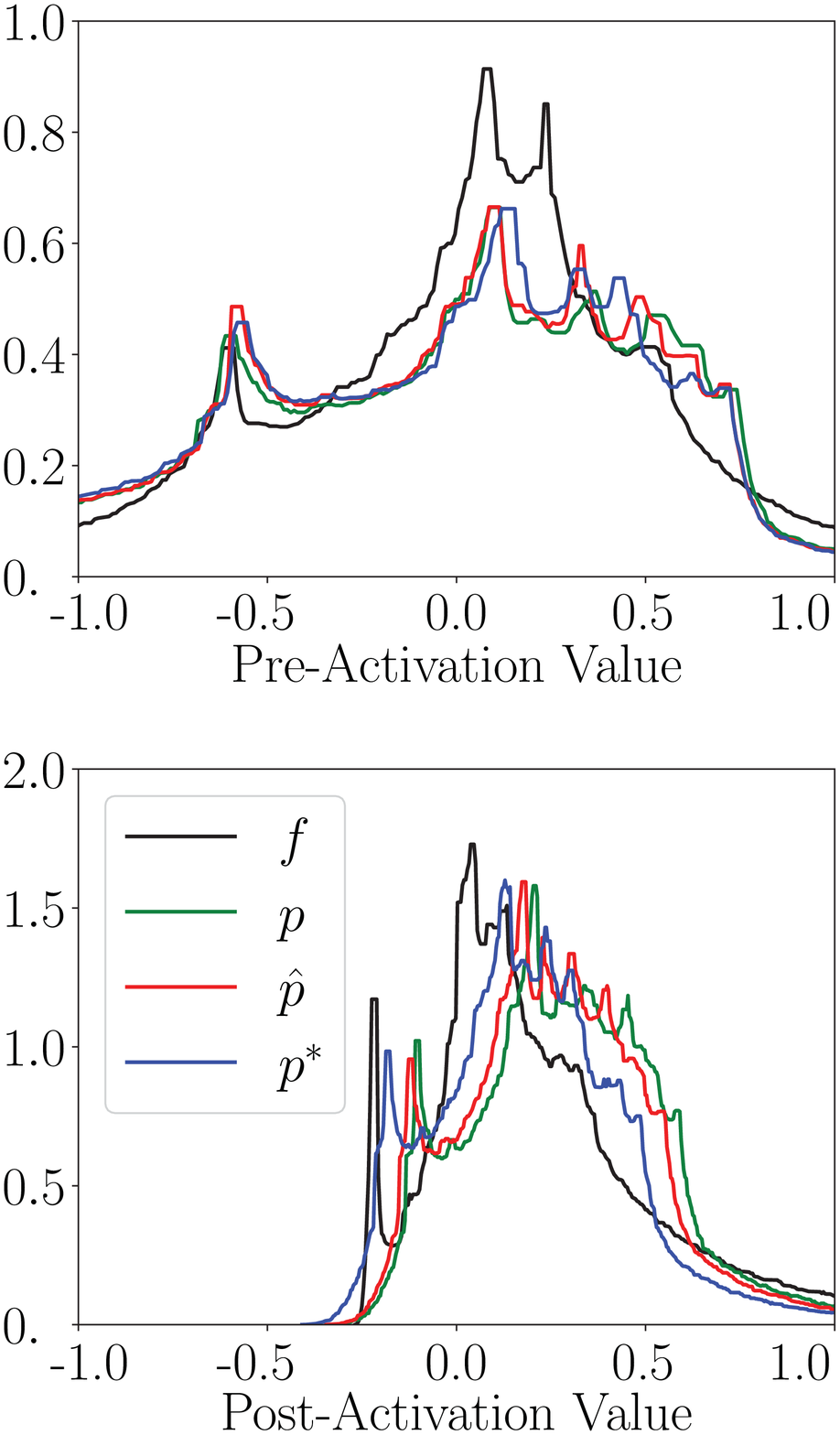}
        \label{fig:hist_bn_conv1}
    }
    \subfigure[Convolution Layer (BN)]{
        \includegraphics[width=1.4in]{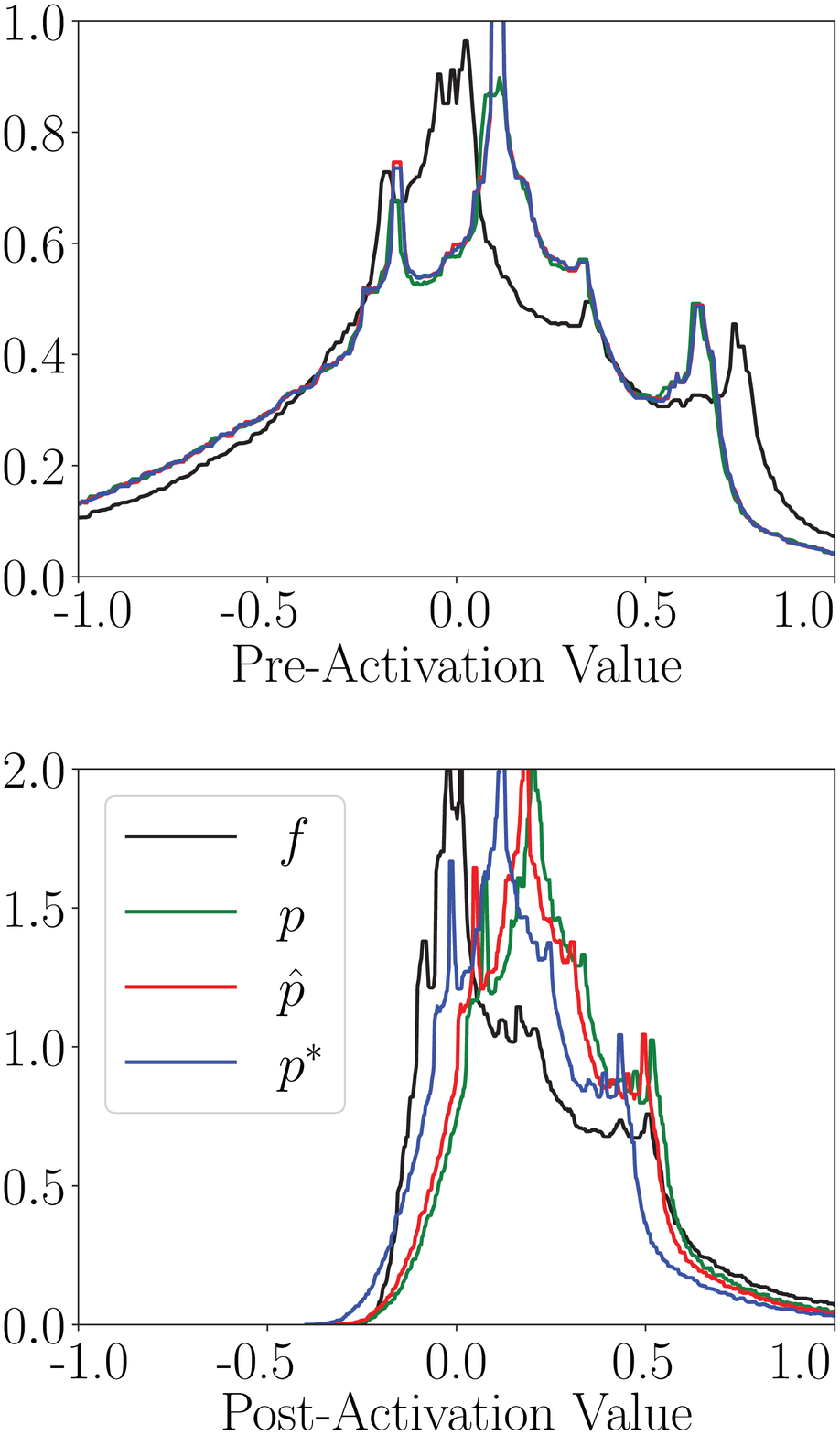}
        \label{fig:hist_no_bn_conv1}
    }
    \subfigure[Dense (fc) Layer]{
        \includegraphics[width=1.4in]{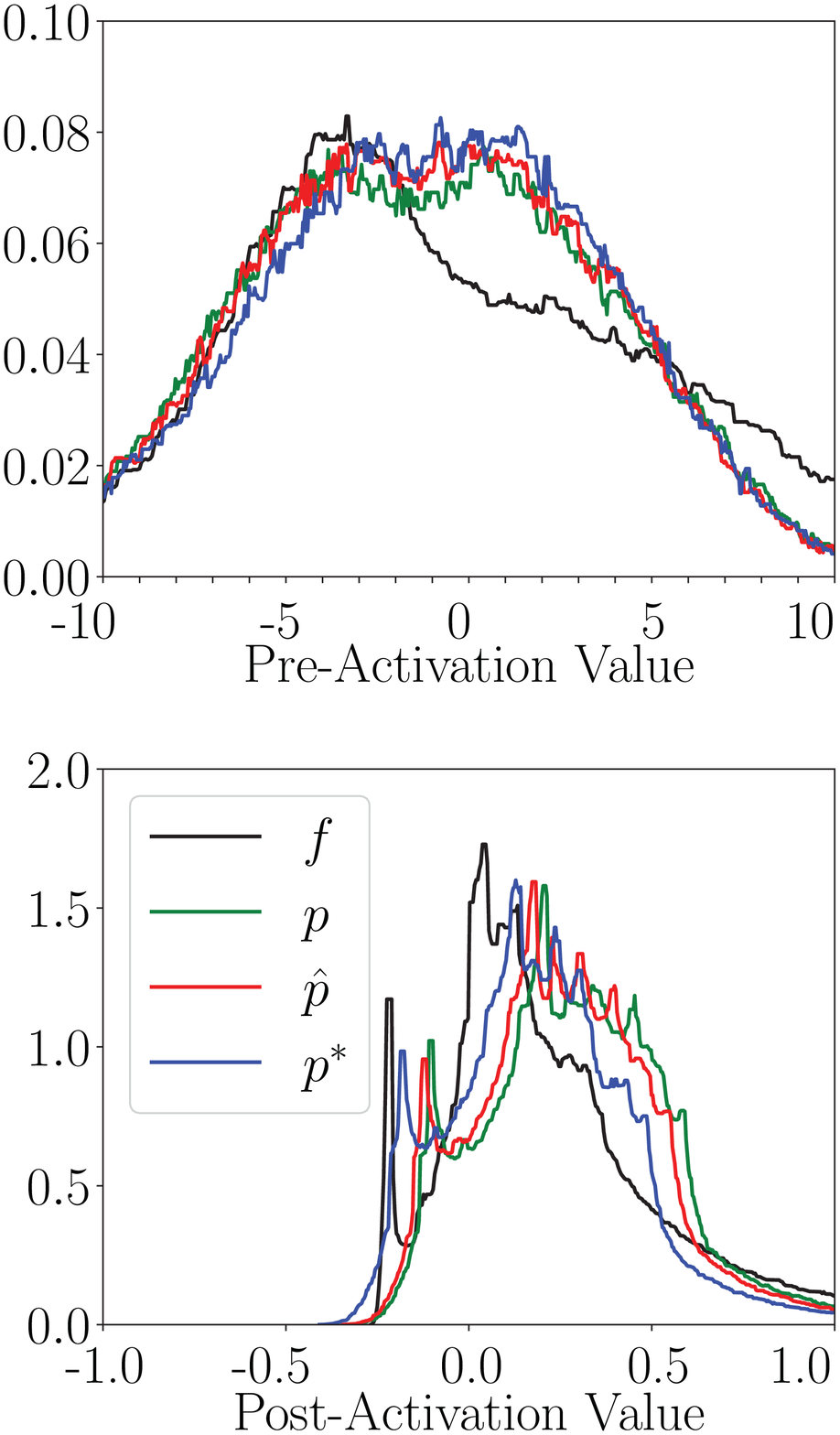}
        \label{fig:hist_bn_fc1}
    }
    \subfigure[Dense (fc) Layer (BN)]{
        \includegraphics[width=1.4in]{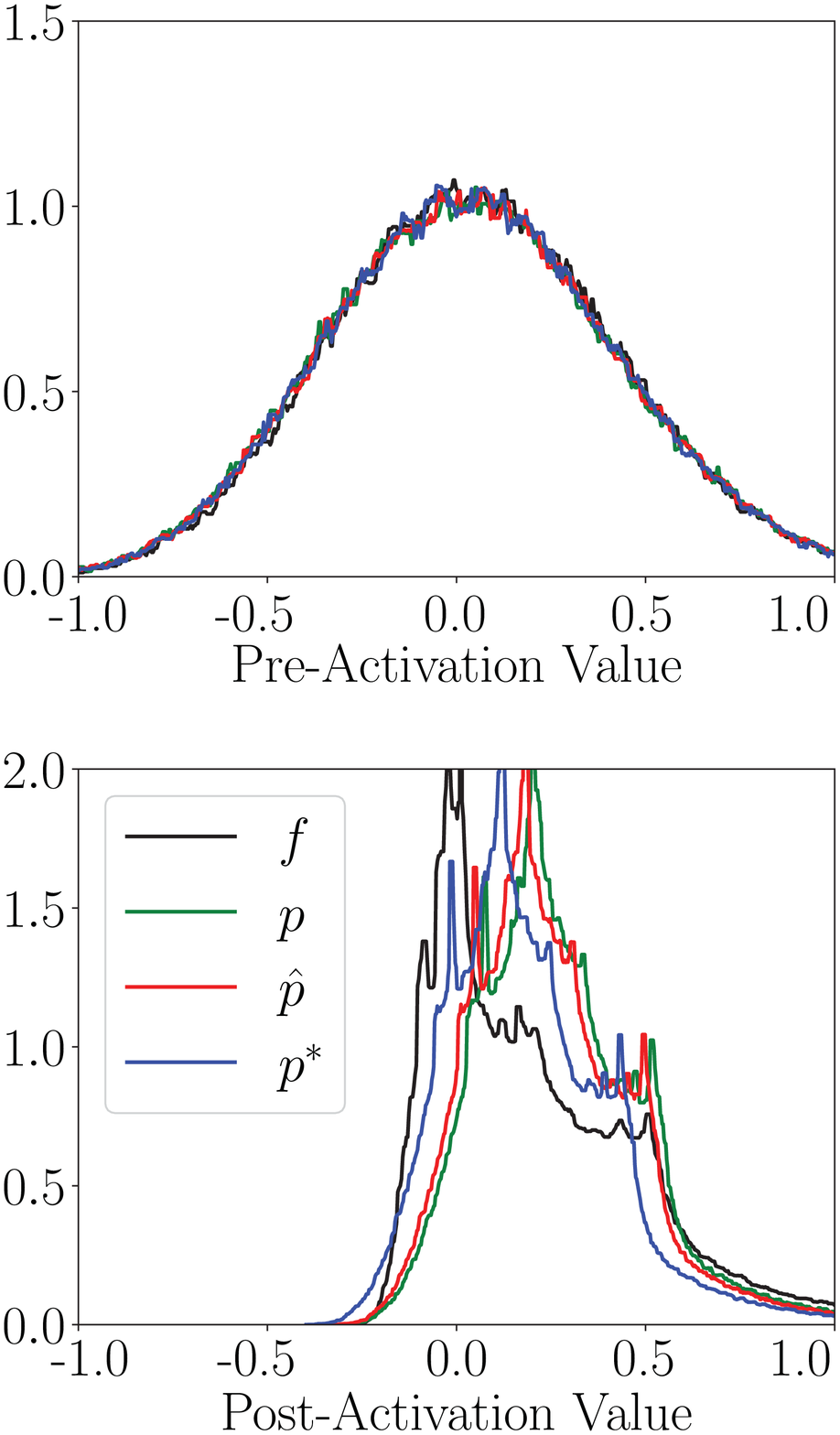}
        \label{fig:hist_no_bn_fc1}
    }
    \caption{Distribution of pre- and post-activation values. (Top) The $x$ axis denotes the pre-activation value. (Bottom) The $x$ axis denotes the post-activation value. (Both) The $y$ axis denotes a normalized frequency. The original activation function is denoted by $f$, the baseline minimax estimate is $p$, the baseline rounded minimax estimate is $\hat{p}$, and our method is $p^*$. BN denotes batch normalization was applied after the convolution but before $f$; this is reflected in the pre-activation value. (Bottom) Values after convolution but before applying $f$.}
    \label{fig:hist}
\end{figure*}

The purpose of the error minimization experiment is to determine which activation function produces the lowest approximation error under our quantization constraints.
We evaluate the effectiveness of multiple approximation schemes including our method.

\subsection{Activation Approximation Accuracy}
We present Table \ref{table:sup_fullactivations} which contains accuracy values for all the layers and all of the activation functions over three trials.
Activation layers we considered include ReLU, square, Swish, and softplus, using the original function, approximated function, and quantized approximation function. 

\begin{table*}[!htbp]
    \centering
    \small
    \begin{tabular}{@{}l|c|c|c|c|c|c|c|c|c|c@{}}
    \toprule
     &  \multicolumn{2}{c|}{Trial 1} & \multicolumn{2}{c|}{Trial 2} & \multicolumn{2}{c|}{Trial 3} & \multicolumn{2}{c|}{Mean} & \multicolumn{2}{c}{Stddev} \\ 
    Activation & Train & Test & Train & Test & Train & Test & Train & Test & Train & Test \\
 \midrule
    Square & 99.80 & 99.08 & 99.81 & 99.14 & 99.8 & 99.29 & 99.80 & 99.17 & 0.01 & 0.11 \\ 
    ReLU & 99.65 & 99.20 & 99.59 & 99.14 & 99.62 & 99.05 & 99.62 & 99.13 & 0.03 & 0.08  \\
    ReLU-approx & 99.57 & 99.07 & 99.60 & 99.14 & 99.58 & 99.07 & 99.58 & 99.09 & 0.02 & 0.04\\ \midrule
    Softplus & 99.42 & 99.17 & 99.37 & 99.06 & 99.41 & 99.05 & 99.4 & 99.09 & 0.03 & 0.07 \\
    Softplus-A & 99.34 & 99.05 & 99.39 & 98.98 & 99.38 & 98.98 & 99.37 & 99.00 & 0.03 & 0.04 \\
    Softplus-AQ & 99.17 & 98.92 & 99.13 & 98.92 & 99.17 & 98.87 & 99.16 & 98.9 & 0.16 & 0.03 \\ \midrule
    Swish & 99.63 & 99.16 & 99.64 & 99.22 & 99.64 & 99.02 & 99.64 & 99.13 & 0.01 & 0.10  \\
    Swish-A & 99.56 & 99.07 & 99.59 & 99.13 & 99.58 & 99.07 & 99.58 & 99.09 & 0.02 & 0.03 \\ 
    Swish-AQ & 99.56 & 99.09 & 99.60 & 99.12 & 99.60 & 99.08 & 99.59 & 99.10 & 0.02 & 0.02 \\
    \bottomrule
    \end{tabular}
    \caption{\textbf{Multiple trials for the activation function ablation study.} Values denote accuracy. Minimax approximation is denoted by \textit{A} and polynomial approximation with quantized coefficients is \textit{AQ} (our method). For each activation function, three models were trained with different random seeds. The mean accuracy and standard deviation are shown.}
    \label{table:sup_fullactivations}
\end{table*}




Figure \ref{fig:approx} shows our approximation methods applied to Swish, ReLU, and softplus. 
The functions are plotted on the top row.
Most approximations are able to fit the original function $f$ within the interval $[-1, 1]$.
The bottom row of Figure \ref{fig:approx} shows the approximation error of $\hat{p}$ and $p^*$ for different pre-activation $x$ values.
Overall, Swish has lower error than ReLU and softplus.
If we can constrain the pre-activation values to fall within the interval, our model will have better approximations.
Conveniently, batch norm transforms the pre-activation values into a normal distribution with zero mean and unit variance \cite{ioffe2015batch} which reduces overall error of the approximation \cite{chabanne2017privacy}.
The shaded area under the curve in Figure \ref{fig:approx} shows the approximation error within the interval $[-1, 1]$.
Swish has lower error than both ReLU and softplus.

In Figure \ref{fig:hist}, we investigate the correctness of our proposed activation approximation method by plotting the pre-activation and post-activation values of different layers for both the regular and approximated Swish functions.
The post-activation graphs in Figure \ref{fig:hist} for Swish show the minimum value between $\approx -0.28$.
We analytically compute the theoretical minimum value for Swish by taking the first order derivative $f'(x) = f(x) + (1 - f(x)) \sigma(x)$.
This gives us the equation $1 + e^{-x} + xe^{-x} = 0$, from which we can derive $x \approx -1.27846$. 
Using $x \approx -1.27846$ to compute $f(x)$, we get an approximate minimum value of $-0.278465$, which corroborates our empirical minimum values shown in Figure \ref{fig:hist}.
We find that this minimum value remains consistent for the approximated Swish function as well, validating the correctness of our approximation method. 

\subsection{Detailed Breakdown of Homomorphic Operations}\label{sup:hops}

\begin{table}[!htbp]
    \centering
    \small
    \begin{tabular}{@{}l|c|c|c|c|c@{}}
    \toprule
    Layer & HOPs & \pbox{20cm}{PT-CT \\ Adds} & \pbox{20cm}{CT-CT \\ Adds} & \pbox{20cm}{PT-CT \\ Mults} & \pbox{20cm}{CT-CT \\ Mults} \\ \midrule
    Conv-1 & 42,757 & 845 & 20,956 & 20,956 & --- \\
    Act-1 & 845 & --- & --- & --- & 845 \\
    Pool-1 &  6,845 & --- & 6,845 & --- & --- \\
    Conv-2 &  309,950 & 1,250 & 154,350 & 154,350 & --- \\
    Pool-2 &  8,450 & --- & 8,450 & --- & --- \\
    FC-1 &  241,192 & 100 & 120,546 & 120,546 & --- \\
    Act-2 & 100 & --- & --- & --- & 100 \\
    FC-2 & 1990 & 10 & 990 & 990 & --- \\  \midrule
    Total &  612,129 & 2,205 & 312,137 & 296,842 & 945 \\ \bottomrule
    \end{tabular}
    \caption{CryptoNets HOPs. More detailed breakdown of HOPs for each layer. Plaintext is denoted by PT and ciphertext is denoted by CT. \textit{Adds} and \textit{mults} refer to the number of homomorphic addition and multiplication operations, respectively. Dashes indicate zero operations. FC refers to the dense (fully-connected) layer.}
        \label{table:sup_hops_cryptonets}
\end{table}

\begin{table}[!htbp]
    \centering
    \small
    \begin{tabular}{@{}l|c|c|c|c|c@{}}
    \toprule
    Layer & HOPs & \pbox{20cm}{PT-CT \\ Adds} & \pbox{20cm}{CT-CT \\ Adds} & \pbox{20cm}{PT-CT \\ Mults} & \pbox{20cm}{CT-CT \\ Mults} \\ \midrule
    Conv-1 & 8619 & 1,690 & 3,042 & 3,887 & --- \\ 
    Act-1 & 5,070 & 845 & 1,690 & 1,690 & 845 \\ 
    Pool-1 & 6,845 & --- & 6,845 & --- & --- \\ 
    Conv-2 & 22,950 & 1250 & 10,850 & 10,850 & --- \\ 
    Pool-2 & 8,450 & --- & 8,450 & --- & --- \\ 
    FC-1 & 14,354 & 100 & 7,077 & 7,177 & --- \\ 
    Act-2 & 600 & 100 & 200 & 200 & 100 \\ 
    Fc-2 & 306 & 10 & 148 & 148 & --- \\  \midrule
    Total & 67,194 & 3,995 & 38,302 & 23,952 & 945 \\ \bottomrule
    \end{tabular}
    \caption{Faster CryptoNets HOPs. More detailed breakdown of HOPs for each layer. Plaintext is denoted by PT and ciphertext is denoted by CT. \textit{Adds} and \textit{mults} refer to the number of homomorphic addition and multiplication operations, respectively. Dashes indicate zero operations. FC refers to the dense (fully-connected) layer.}
        \label{table:sup_hops_fastcryptonets}
\end{table}

In Table \ref{table:sup_hops_cryptonets}, we list the HOPs at a more granular level than those presented in Table \ref{table:sota} for CryptoNets.  In Table \ref{table:sup_hops_fastcryptonets}, we list the HOPs for our \textit{Faster CryptoNets} method.  We can see that the number of HOPS is greatly reduced for each layer and for the overall network.




\subsection{Comparison with Prior Work}\label{sec:sota}


\begin{table*}[!htbp]
\centering
\small
\begin{tabular}{l|c|ccc}
    \toprule
    Criteria & Faster CryptoNets & CryptoNets & CryptoDL-1 & CryptoDL-2  \\ \midrule
    PT-CT Adds & 3,995 & \textbf{2,205} & 30,750 & 161,546\\
    CT-CT Adds & \textbf{38,302} & 312,137 & $2.31 \times 10^6$ & $4.61 \times 10^7$  \\
    PT-CT Mults & \textbf{23,952} & 296,842 & $2.31 \times 10^6$ & $4.62 \times 10^7$  \\
    CT-CT Mults & \textbf{945} & \textbf{945} & 1,600 & 64,512 \\
    Total HOPs & \textbf{67,194} & 612,129 & $4.65 \times 10^6$ & $9.27 \times 10^7$ \\  \midrule
    Encrypt+Decrypt Time & \textbf{6.7 sec} & 47.5 sec & 16.7 sec & 16.7 sec \\
    Inference Time & \textbf{39.1 sec} & 249.6 sec & 148.9 sec & 320.0 sec \\ \midrule
    Test Set Accuracy & 98.71 & 98.95 & 98.52 & \textbf{99.52} \\
    Message Size & 411.1 MB & 367.5 MB & \textbf{336.7 MB} & \textbf{336.7 MB} \\ 
    Encryption Scheme & FV-RNS & YASHE & BGV & BGV \\ 
    \bottomrule
\end{tabular}
\vspace{-1mm}
\caption{\textbf{Comparison of State-of-the-Art Methods (Encrypted).} Plaintext is denoted by PT and ciphertext is denoted by CT. \textit{Adds} and \textit{mults} refer to the number of homomorphic addition and multiplication operations, respectively. The total number of homomorphic operations is denoted by HOPs. Message size is the size of a single encrypted image. 
\textit{Faster CryptoNets} uses Swish-AQ while  CryptoNets uses $f(x)=x^2$ as the activation function. References: CryptoNets \cite{gilad2016cryptonets}, CryptoDL \cite{hesamifard2017cryptodl}.}
\label{table:sota}
\end{table*}

The target use case of our work is inference on a single encrypted image (Figure \ref{fig:pull}).
We believe this approach is more analogous to practical use cases, where the third-party host runs asynchronous inference for individual users.
Additionally, \cite{Migliore:2017:HAH:3145508.3126558} suggests that there are very significant drawbacks to batching, including having to select more numerous and restricted NTT points, forcing specific computations away from NTT, and adding large computational cost.
Works focusing on accelerating neural networks neglect batching for similar reasons as ours (\cite{sanyal2018tapas} does not use batching, and \cite{florian2017fastdiscretize}. uses batching to compress messages but not to improve throughput). Works that do batch inputs use schemes not very efficient in practice (discussed in \cite{Migliore:2017:HAH:3145508.3126558}.) and do not report the batching cost. A thorough performance analysis of batching binary vs scalar messages across different libraries is beyond the scope of our paper but would be a great direction for future work. 
As such, we do not implement ciphertext batching techniques in this paper, although we find it worth noting that our technique does not preclude the use of batching techniques. \cite{Migliore:2017:HAH:3145508.3126558} introduces the Karatsuba algorithm which supports batching with binary encoding, preserving the benefits from our method.

We refer (Table \ref{table:sota}) for accuracy and runtime results.
The test set accuracy of our original model is 99.12\%, and is slightly reduced to 98.71\% after pruning and quantization.
Evaluation of network layers in \textit{Faster CryptoNets} takes 39.1 seconds for one input, compared to 249.6 seconds for CryptoNet.  We achieve a $6.4\times$ improvement in wall-clock time while maintaining accuracy comparable to that of CryptoNets, which achieved 98.95\% test accuracy.  We also find that our method achieves $9.1 \times$ fewer HOPs, a larger improvement than raw wall-clock time suggests. 
In \textit{Faster CryptoNets}, encoding/encryption takes 6.63 seconds, while decryption of the final layer's output takes 0.02 seconds.
CryptoNets takes 44.5 seconds for encoding/encryption, and 3 seconds for decryption.
Our method is $6.7\times$ and $150\times$ faster for these operations, respectively.

MNIST images are $28 \times 28$ pixels.
Each ciphertext consists of 2 polynomials resulting in 65,544 integers (64-bit). 
Therefore, our message consists of $28 \times 28 \times 65544 \times 8$ bytes, or 411.1 MB. The output of the network consists of the 10 outputs of the final dense layer, which gives us a result consisting of $10 \times 65544 \times 8$ bytes, or 5.24 MB. In CryptoNets, the authors' encryption scheme results in each image consuming 367.5 MB in encrypted form.
Our scheme results in comparable message sizes to previous work.

\subsection{Ablation Studies}\label{sec:ablation}

\begin{table}[!htbp]
    \centering
    \footnotesize
    \begin{tabular}{@{}l|cc|cc|cc@{}}
    \toprule
     & \multicolumn{2}{c|}{Faster CryptoNets} & \multicolumn{2}{c|}{CryptoNets} & \multicolumn{2}{c}{Relative} \\ \midrule
    Layer & Time & HOPs & Time & HOPs & Time & HOPs \\
    \midrule
    Conv-1 & \textbf{3.9} & \textbf{8,619} & 30.0 & 42K& $7.7 \times$ & $4.9\times$ \\
    Act-1 & \textbf{23.4} & 5,070 & 81.0 & 845 & $3.5 \times$ & $0.2 \times$ \\
    Mid & \textbf{9.1} & \textbf{53K} & 127.0 & 566K & $14 \times$ & $11 \times$ \\
    Act-2 & \textbf{2.7} & 600 & 10.0 & 100 & $3.7 \times$ & $0.2\times$ \\
    FC-2 & \textbf{0.1} & \textbf{306} & 1.6 & 1,990 & $16\times$ & $6.5\times$ \\ \midrule
    Total & \textbf{39.1} & \textbf{67K} & 249.6 & 612K & $6.4 \times$ & $9.1\times$ \\ \bottomrule
    \end{tabular}
    \caption{\textbf{Layer-Wise Analysis (Encrypted).} Wall-clock time (seconds) and HOPs required for inference on a single encrypted image. K denotes thousands. \textit{Act} refers to the activation function. \textit{Mid} denotes a combination of pool1, conv2, pool2, and fc1, as reported by \cite{gilad2016cryptonets}. 
    }
        \vskip -0.1in
    \label{table:layerwise}
\end{table}

\textit{Faster CryptoNets} differs from the CryptoNets model in that we use the Swish activation instead of the square function.
While both methods use a 2\textsuperscript{nd} degree polynomial of the form $p(x) = a_0 + a_1 x + a_2 x$, our approximations use $a_0, a_1 \neq 0$ for increased expressivity whereas the square function set $a_0 = a_1 = 0$, resulting in fewer HOPs for the square function.
This is shown in Table \ref{table:layerwise} in the rows Act-1 and Act-2.
Despite our method requiring more HOPs for Act-1 and Act-2, we still achieve a faster inference time than CryptoNets.
At a per-layer level, our method yields up to $16\times$ and $11 \times$ improvements for wall-clock and HOPs, respectively.

We compare the performance of different activation functions when approximated with our proposed polynomial approximation and quantization (AQ) method.
For MNIST, Swish-AQ produced a test accuracy of 99.10\%, while ReLU-AQ achieves an equivalent test accuracy of 99.09\%.  
We note the similarity of ReLU and Swish.
This finding is corroborated by the similarity of the approximations in Figure \ref{fig:approx}.
The polynomial coefficients we calculate for the ReLU and Swish approximations turn out to be the same, except for a constant factor.


We evaluate the inference quality and runtime of pruning and quantization separately.
Pruning produced a test set accuracy of 98.73\% and inference time of 104.7 seconds.
Quantization produced 99.06\% and 162.5 seconds.
When combined, pruning and quantization produced 98.71\% and 45.7 seconds.
We record the accuracies during each INQ step for both a non-pruned network in Table \ref{table:sup_quantize1} and a DNS pruned network in Table \ref{table:sup_quantize2}.
The accuracy is largely preserved as the network is successively quantized, demonstrating consistent preservation.
Overall, accuracy was preserved, or slightly improved in the case of quantization. 
\begin{table}[!htbp]
    \centering
    \small
    \begin{tabular}{@{}c|c|c|c@{}}
    \toprule
    INQ Step &  Partition & Quantized\% & Accuracy \\ \midrule
     1 & 0.7 & 30\% & 99.00 \\
    2 & 0.4 & 60\% & 99.02 \\ 
    3 & 0.2 & 80\% & 98.99 \\ 
    4 & 0.0 & 100\% & 99.06 \\ 
     \bottomrule
    \end{tabular}
    \caption{INQ-only quantization schedule. Accuracies collected for each Incremental Network Quantization (INQ) step are reported in the Accuracy column. In each step, a progressively larger set of the weights are partitioned and quantized, as reported in columns 2 and 3.}
    \label{table:sup_quantize1}
\end{table}

\begin{table}[!htbp]
    \centering
    \small
    \begin{tabular}{@{}c|c|c@{}}
    \toprule
    INQ Step & Partition  & Accuracy \\ \midrule
     1 & 0.98 &  98.74\\
    2 & 0.96 &  98.78\\ 
    3 & 0.94 &  98.69\\ 
    4 & 0.92 &  98.68\\ 
    5 & 0.90 &  98.69\\
    6 & 0.88 &  98.79\\ 
    7 & 0.86 &  98.68\\ 
    8 & 0.00 &  98.71 \\ 
     \bottomrule
    \end{tabular}
    \caption{DNS+INQ quantization schedule. Similar analysis is performed on a Dynamic Network Surgery (DNS) pruned network.  Accuracies collected for each INQ step are reported in the Accuracy column, as well as the weight partitioning in the Partition column.}
    \label{table:sup_quantize2}
\end{table}

\section{Experimental Correctness}
We make sure our parameters are selected properly so that our decrypted outputs are correct.  We run encrypted inference on the 10,000 image MNIST test-set, and find no accuracy loss from our method's plaintext results (98.71\%).  We also find an precision error of around 0.05\% when comparing the plaintext and decrypted outputs.  Upon further examination, we found that his error is introduced during the floating point to fixed point conversion prior to the encoding scheme, and that this error does not effect the accuracy of our model.

\section{Scaling up}\label{sec:scaling up}
To evaluate how well our method works in real-world settings, we implement our techniques on larger datasets.  First, we focus on CIFAR-10 as a larger practical image classification task. Next, we consider diabetic retinopathy dataset as a real-world medical imaging use-case where very deep neural networks would be used in practice.
For both experiments, we upgrade our machines to n1-megamem-96 instances offered by the Google Cloud Platform, which each have 96 Intel Skylake 2.0 GHz vCPUs and 1433.6 GB RAM. 

\subsection{FV-RNS Parameters}
We use a ring dimension $n = 8192$ with fifteen plaintext moduli $t^{(j)}$: 40961, 65537, 114689, 147457, 188417, 270337, 286721, 319489, 417793, 557057, 638977, 737281, 778241, 786433, 925697. The values of the coefficient moduli $q^{(j)}$ are selected to provide 128-bit security, such that $\log q^{(j)} = 219$. Furthermore, each coefficient modulus $q^{(j)}$ is decomposed into four 64-bit moduli for efficient use of the RNS variant of the FV encryption scheme.

\section{CIFAR-10}
MNIST is a relatively easy dataset, with simple machine learning algorithms like linear regression or KNN producing high accuracy results \cite{toghi2018mnistknn}. CIFAR-10 \cite{cifar10} is a more complicated task where CNN's perform notably better than other methods.
We evaluated the CIFAR-10 performance of our method on the model used in CryptoDL \cite{hesamifard2017cryptodl}, consisting of eight convolutional layers, which from now on we will denote as CNN-8.

\begin{table}[!htbp]
    \centering
    \small
    \begin{tabular}{@{}l|c|c|c|c@{}}
    \toprule
    Activation & \pbox{20cm}{CIFAR-10 Train Acc.} & \pbox{20cm}{CIFAR-10 Test Acc.} \\ \midrule
    ReLU & $\mathbf{93.10}$ & $\mathbf{86.76}$  \\
    Square & $59.97$ & $59.88$ \\
    Softplus & $67.96$ & $67.94$ \\ 
    Swish  & $91.55$ &	$86.24$\\   \midrule
    ReLU-A  & $\mathbf{77.95}$ & $\mathbf{75.99}$ \\
    Softplus-A & $75.11$ & $73.57$ \\ 
    Swish-A & $77.30$ & $75.41$ \\ \midrule
    ReLU-AQ & $77.95$	& $\mathbf{75.99}$\\
    Softplus-AQ & $72.67$ & $71.58$ \\
    Swish-AQ & $\mathbf{78.20}$ & $75.66$ \\ \bottomrule
    \end{tabular}
    \caption{Approximation results. Minimax approximation is denoted by \textit{A} and polynomial approximation with quantized coefficients is \textit{AQ} (our method). The training accuracy and test accuracy are shown for CIFAR-10.}
    \label{table:activation_times_cifar}
\end{table}

\begin{table}[!htbp]
    \centering
    \small
    \begin{tabular}{@{}l|c|c|c|c@{}}
    \toprule
    Layer  & \pbox{20cm}{PT-CT \\ Adds} & \pbox{20cm}{CT-CT \\ Adds} & \pbox{20cm}{PT-CT \\ Mults} & \pbox{20cm}{CT-CT \\ Mults} \\ \midrule
    Conv-1	& 36,864	& 460,800 &	479,232 &	0 \\
    Conv-2 &	36,864 &	13,294,908 &	13,313,340 &	--- \\
    Activ-1 &	18,432 &	36,864 &	55,296 &	36,864 \\
    Pool-1  &	--- &	18,432 &	--- &	--- \\
    Conv-3 &	18,432 &	5,968,347 &	5,977,563 &	0 \\
    Conv-4 & 18,432 &	11,931,014 &	11,940,230 &	--- \\
    Activ-2 &	9216 &	18,432 &	27,648 &	18,432 \\
    Pool-2 & 0 &	9216 &	--- &	--- \\
    Conv-5 &	9216  &	4,713,389 &	4,717,997 &	0 \\
    Conv-6 & 9,216	 & 9,421,992 &	9,426,600 &	0 \\
    Activ-3 &	4,608 &	9,216 &	13,824 &	9,216 \\
    Pool-3 &	--- &	4,608 &	--- &	--- \\
    FC-1 &	256 &	29,4644 &	294,644 &	--- \\
    FC-2 & 10 &	2560 &	2,560 &	0 \\  \midrule
    Total &	161,546 &	46,184,422 &	46,248,934 &	64,512 \\ \bottomrule
    \end{tabular}
    \caption{CNN-8 HOPs. More detailed breakdown of HOPS for each layer. Plaintext is denoted by PT and ciphertext is denoted by CT. \textit{Adds} and \textit{mults} refer to the number of homomorphic addition and multiplication operations, respectively. Dashes indicate zero operations. FC refers to the dense (fully-connected) layer.}
    \label{table:sup_hops_cryptodl2}
\end{table}

\subsection{Activation Comparisons}
We present Table \ref{table:activation_times_cifar} which contains accuracy values for all the layers and all of the activation functions.
Activation layers we considered include ReLU, square, Swish, and softplus, using the original function, approximated function, and quantized approximation function. 
In Table \ref{table:activation_times_cifar}, we can see that training this model with the square activation function resulted in significantly worse test accuracy (59.88\%) compared to training this model with the ReLU activation function (86.76\%), confirming the theoretical loss of accuracy from instability of the square function for deeper neural networks.
Furthermore, we find that ReLU-AQ and Swish-AQ offer comparable levels of performance (77.95\% and 78.20\% training accuracy, 75.99\% and 75.66\% test accuracy), while significantly improving on the accuracy results that were achieved with the square activation function.

\subsection{Pruning and Quantization}
The pruning and quantization procedure results in a model with slightly improved accuracy (76.72\%) that requires an order of magnitude fewer HOPS for inference ($6.12\times10^8$ HOPs vs. $6.47\times10^9$ HOPs for the baseline method).
The inference time for the model was 22,372 seconds with our method.

\subsection{Message Size}
The message size for the input image is $32 \times 32 \times 3 \times 65544 \times 8$ bytes, or 1,610.8 MB.

\section{Medical Imaging}

\begin{figure*}[!htbp]
\centering
\includegraphics[width=5.5in]{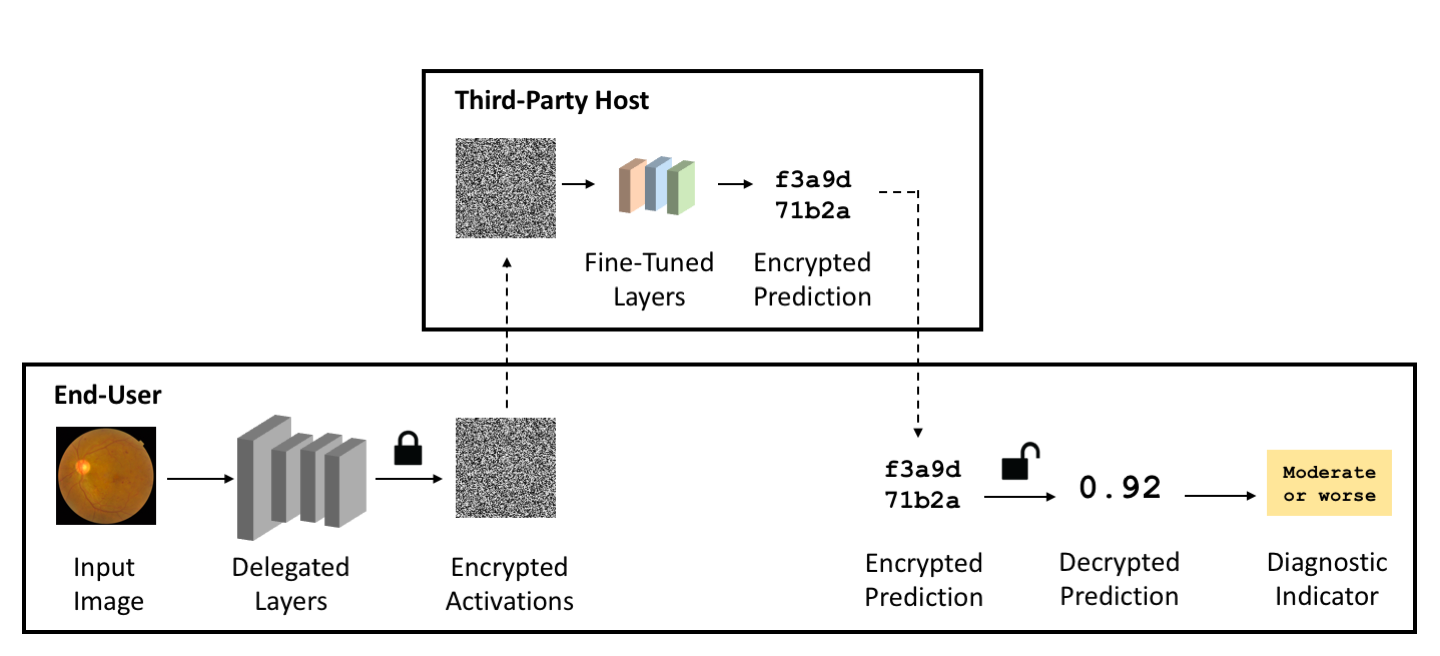}
\caption{Illustration of our proposed method for encrypted inference on retinal fundus images. The cloud provider delegates the computation of the pretrained layers of the neural network to the end-user and evaluates the task-specific layers using the homomorphic encryption scheme. The output of the network is returned in encrypted form to the end-user, who can decrypt to determine the prediction of the severity of diabetic retinopathy.}
\label{fig:method}
\end{figure*}

\begin{table}[!htbp]
    \centering
    \footnotesize
    \begin{tabular}{|c|c|c|} \hline
    Model & Layers Retrained & Test Accuracy \\ \hline
    CNN-8 & All Layers & 63.23  \\ 
    DFE-RN-50 & Top Block & 69.89  \\
    DP-DFE-RN-50 & All layers (with DP) & 76.47  \\
    ResNet-50 & All Layers & $\mathbf{80.61}$ \\  \hline
    \end{tabular}
    \caption{Accuracies and layers trained of each model. RN denotes ResNet, DFE denotes delegated feature extraction, and DP denotes differentially private.  A ResNet-50 model is trained in a standard setting for benchmarking. We observe that DFE-RN-50 has significantly higher accuracy than the baseline CNN model and that DP-DFE-RN-50 further increases the test accuracy close to a plain ResNet-50 model.  }
    \label{tab:accuracy_results}
\end{table}

\begin{table}[!htbp]
    \centering
    \footnotesize
    \begin{tabular}{|c|c|c|c|c|} \hline
    Model & Accuracy & HOPs & Runtime (s) \\ \hline
    CNN-8 (sparse) & 63.23 & 1.33E8 & 3325  \\ 
    DP-DFE-RN-50  & 76.04 & 3.95E8 & $12493$ \\
    DP-DFE-RN-50 (sparse) & $\mathbf{76.47}$ & $\mathbf{4.23E7}$ & $\mathbf{1590}$ \\ 
    \hline
    \end{tabular}
    \caption{Comparison of performance metrics for transfer learning vs. fully-trained models. We show that our privacy-safe delegated feature extraction model results in both higher accuracy and fewer HOPs/runtime, and also show that our sparsity techniques maintains high accuracy.}
    \label{tab:performance_results}
\end{table}

\begin{table}[!htbp]
    \centering
    \begin{tabular}{|c|c|c|c|c|} \hline
    Method & Acc. & HOPs & \pbox{20cm}{Inference \\ Time (s)} & Speedup \\ \hline
    Original & 70.47 & $3.95\times10^8$ & 12493 & -- \\ 
    Pruned & \textbf{70.98} & $\mathbf{4.23\times10^7}$ & 1924 & 6.4x \\
    \pbox{20cm}{Pruned/Quantized} & 70.55 & $\mathbf{4.23\times10^7}$  & \textbf{1590} & \textbf{7.8x} \\ \hline
    \end{tabular}
    \caption{Ablation study of methods for improving performance of DFE-ResNet-152}
    \label{tab:ablation_study}
\end{table}


\begin{figure}[!htbp]
\centering
\includegraphics[width=1.5in]{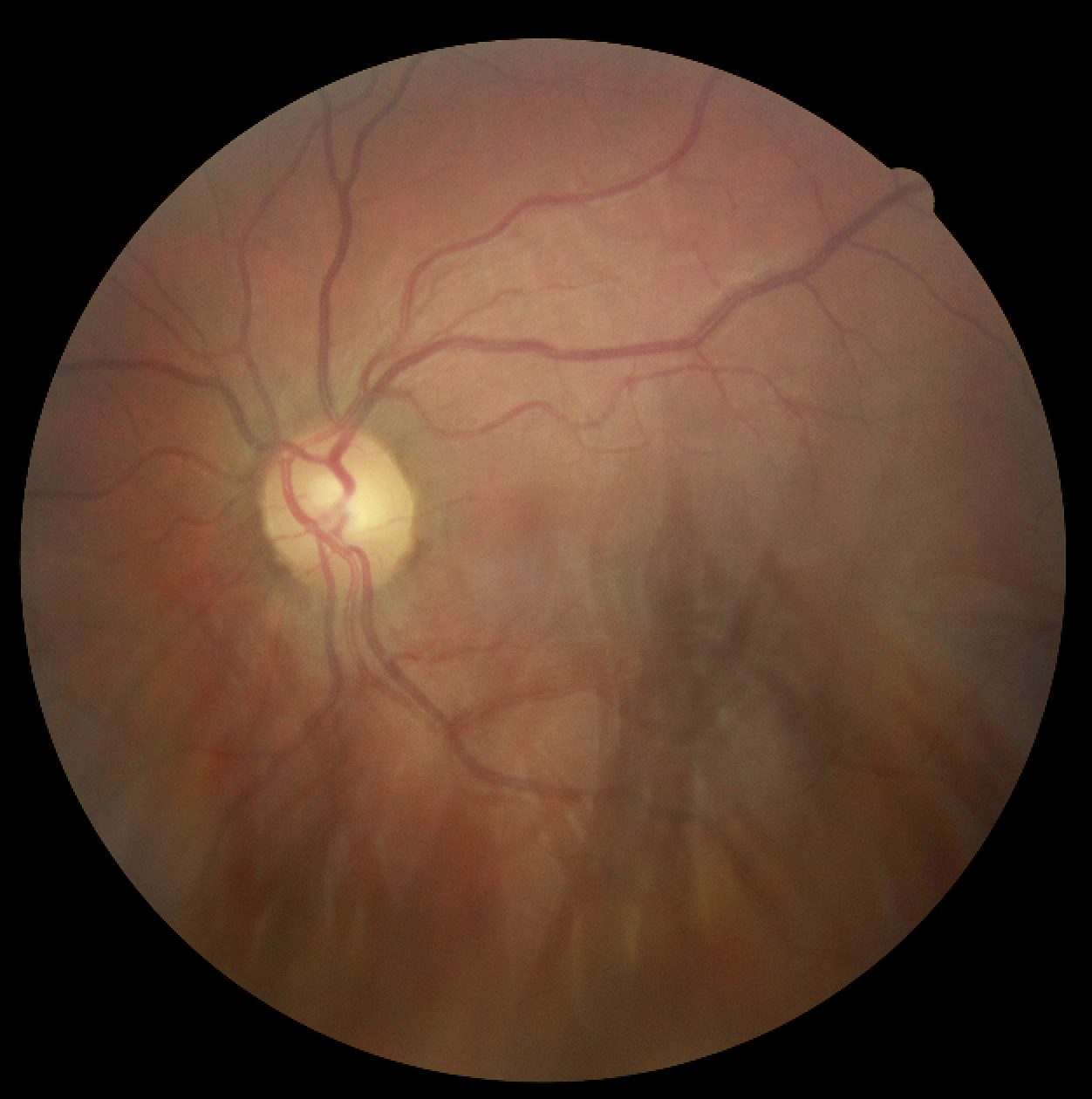}
\includegraphics[width=1.5in]{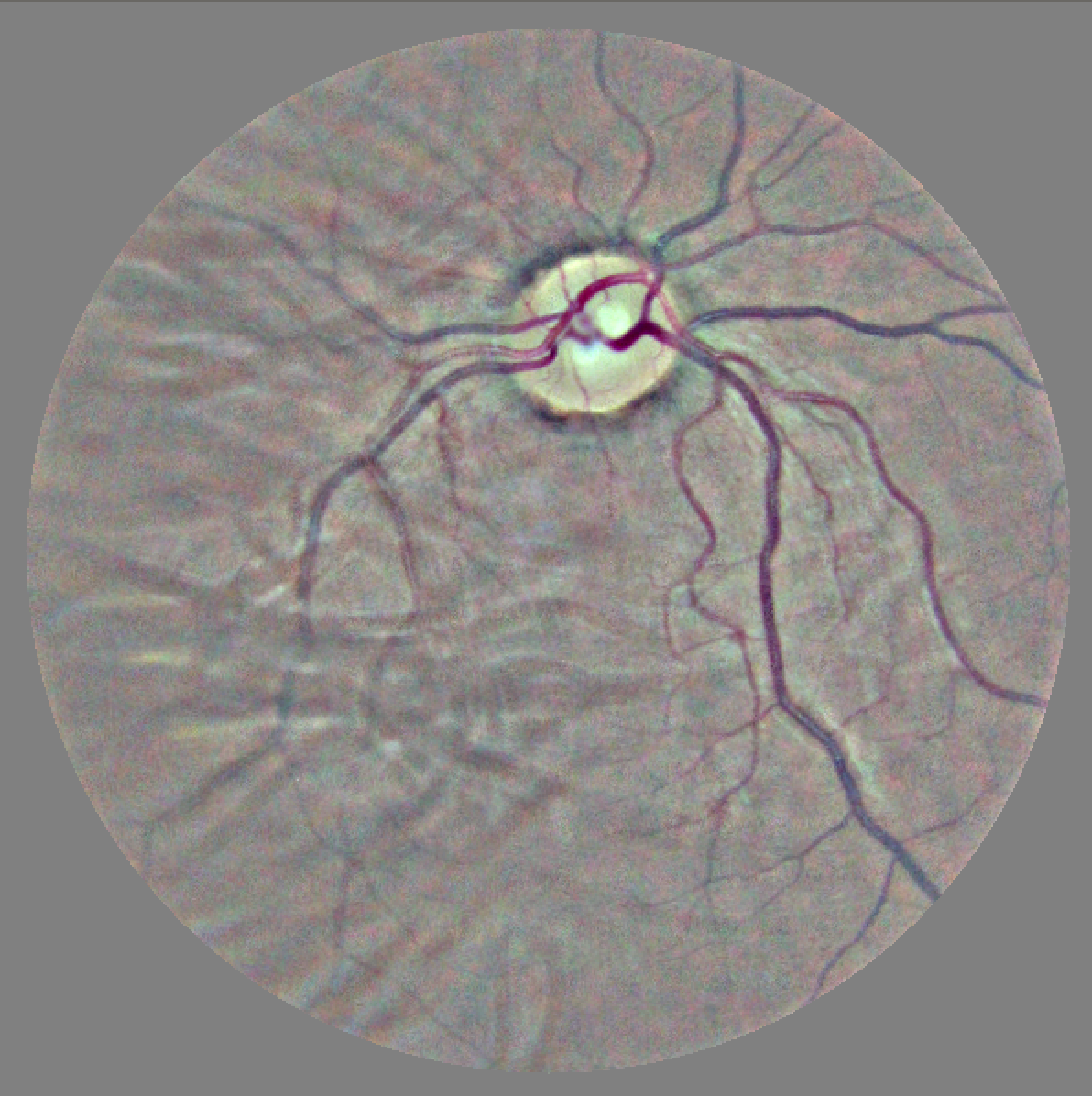}
\caption{Illustration of retinal fundus image, graded with `none' rating for diabetic retinopathy. The right image shows the result of the procedure for preprocessing.}
\label{fig:data_example}
\end{figure}

A significant limiting factor in levelled encryption schemes used with neural networks is multiplicative depth, where only set amount of HE operations can be performed sequentially.  
Increasing the multiplicative depth by choosing larger parameters  incurs prohibitive cost, limiting us to neural networks with three activation functions with our current settings.
However, state-of-the-art real world applications for deep learning like medical imaging applications commonly use modern very deep neural networks. 
To mitigate this issue, we propose the use of models trained with transfer learning, where the computations involved in the pretrained layers of the model can be delegated to the client, and encryption is applied only for the evaluation of the fine-tuned layers on the server. 
Using this technique, which we call \textit{Delegated Feature Extraction} (DFE), as well as \textit{Faster CryptoNet} optimizations to speed up the computation, we achieve practical runtimes for large input sizes.  An illustration of our technique is provided in Figure \ref{fig:method}.
We show how this technique can be improved with private training, demonstrating a viable framework where private, efficient, and powerful machine learning services can be provided.


\subsection{Data}
We chose the diabetic retinopathy dataset introduced by \cite{gulshan2016development} both for its clinical impact and the privacy-sensitive nature of retinal data.  The dataset consists of macula-centered retinal fundus images primarily sourced from EyePACS and was graded by 54 opthalmologists or opthalmologist trainees using the International Clinical Diabetic Retinopathy scale \cite{severityscale} into `none', `mild', `moderate', `severe', or `proliferative' ratings for the severity of the condition.

We were able to obtain a subset of around 35,126 images of the dataset, with a label distribution of 25,810 `none', 2,443 `mild', 5,292 `moderate', 873 `severe', and 708 `proliferative' diagnoses.  Scans from both the left and right eye were sourced from each patient.  To compare our results to the replication study performed by \cite{replicationstudy}, we group the `none' and `mild' labels to a `0' label and the `moderate', `severe', and `proliferative' labels to a `1' label to reframe our problem into a binary classification task.  We randomly subsampled our dataset to get an even split between our two labels, and following the guidelines recommended by \cite{gulshan2016development} we use an 80-20 split for training and test data.

Before using the retinal images with our network, we perform some preprocessing on the raw images. The scans are scaled to $224 \times 224 \times 3$, the standard ImageNet input size, with cropping performed using edge detection to reframe the images.  To normalize the colors and lighting, each image is subtracted by the local average color of each image, after which the local average is mapped to grayscale.  Random rotation is performed on the image to make the model invariant to left/right eye positioning and for general augmentation.  Samples of the data we use are provided in Figure \ref{fig:data_example}.

\subsection{Transfer Learning}
Transfer learning is a useful way to learn an accurate model with limited data \cite{bengio2012deep}, and requires retraining (fine-tuning) of only a few of the layers of the network rather than training a full network from scratch \cite{hinton2006reducing}. The base layers are commonly trained on ImageNet \cite{deng2009imagenet, russakovsky2015imagenet}, and the final layers are retrained on a specific proprietary dataset. This practice is common in the healthcare setting, where large datasets are expensive to acquire and transfer learning can simplify and/or improve the training of the model, especially when using Machine Learning as a Service (MLaaS) offerings from cloud computing providers to expedite development of the application \cite{tajbakhsh2016convolutional, greenspan2016guest}. As we describe in more detail in our methods section, we can leverage transfer learning such that only a small number of fine-tuned layers of the network require evaluation under the encryption scheme, while the generic feature extraction of the base network layers is delegated to the client.

\subsection{Network Architectures}
We first implement a baseline model to train on the retinal dataset. Our baseline model (CNN-8) resembles the CNN network architecture presented in \cite{hesamifard2017cryptodl} for CIFAR-10, but is designed to support the larger input image sizes (scaled from $32\times32\times3$ to $224\times224\times3$ input). It leverages an identical multiplicative depth budget to our transfer learning models in realizing the privacy guarantees. In particular, it contains eight convolutional layers and three approximate activation functions. 

Modern deep neural networks like ResNet-152 \cite{he2015resnet} and Inception-v3 \cite{szegedy2015inception} can contain hundred of layers. The multiplicative depth of the fully-trained ResNet-152 or the Inception-v3 in \cite{gulshan2016development} is at least an order of magnitude greater and would have a prohibitively large runtime in the encrypted setting. Additionally, there exist challenges in achieving strong accuracy when training all layers of a deep neural network with approximate activation functions. Our proposed model (DFE-ResNet-152) only requires retraining of the top block of the ResNet-152 network, which contains three convolutional layers and three activation functions.  For this top block, we replace the ReLU activation functions in the top block with our approximate activation function, and we use the rest of the ImageNet-pretrained model as a delegated feature extractor on the client.

\subsection{Model Adaptations}

We use the approximation of Swish \cite{ramachandran2017swish} given by $p^*(x) = 2^{-3}x^2 + 2^{-1}x + 2^{-4}$ as derived in previous sections. This approximation is required for only the activation functions in the retrained block, which reduces difficulties with convergence in training.

Since the encryption scheme only supports addition and multiplication, some minor modifications are required to support average pooling. While other work has used scaled average pooling, in this work we encode the reciprocal of the size of the pooling window. Furthermore, to support the addition operation in the residual block, the scaling factor must be encoded as well, to scale the encrypted input image for the addition operation.

\subsection{Client-Server Interaction}

The client uses a standard deep learning framework to evaluate the base network layers on a single RGB retinal fundus image. Each element of the activation volume of the final base layer can be converted to a fixed-point value and encoded using the integer encoder described above. Each value is encrypted on the client and transmitted to the server. The server returns encrypted output values, which the client decrypts, converts to floating point, and applies the sigmoid function to learn the final predictions of the model. Given the support for deep learning operations in major mobile platforms, the client could even be the patient's own mobile device, allowing for direct service models in developing countries.

\subsection{Evaluation Metrics}\label{eval_metrics}

The natural alternative to our proposed technique is to have the entire fully-trained model stored on the server and for clients to transmit their encrypted images for diagnosis. 
We want to demonstrate that our proposed technique of using fine-tuned layers of a much deeper model has higher accuracy, greater performance, and a smaller message size, without compromising any security or privacy guarantees. Our primary points of comparison will be between a standard model that is fully-trained on our dataset and a transfer learning model that uses pretrained ImageNet weights and is fine-tuned on our dataset. 

We will need to consider the multiplicative depth of the models, which corresponds to the length of the deepest path of ciphertext multiplications through the network. We keep the multiplicative depth fixed between the two methods to enable a fair comparison. We will also analyze the count of homomorphic operations (HOPs), which serves as a hardware-independent and implementation-agnostic measure of the complexity of evaluation.

\subsection{DP-SGD}

Once concern of the earlier approach is that the users are limited to fine-tuning only a few layers of a network. 
However, training the entire model and release the upper portion of the network as a feature extractor could lead to a privacy risk as the feature extractor could potentially leak sensitive information.
We explore the use of private training techniques to fine-tune the feature extractor further to improve accuracy while preserving end-to-end privacy.  

Differential privacy is a privacy construct which guarantees that an individual will not change the overall statistics of the population \cite{dwork2008differential}. Formally, it is defined that an algorithm $M$ and dataset $D$ are $(\epsilon,\delta)$ private if $P(M(x \in D) \in S)\leq e^\epsilon P(M(x \in D')\in S) + \delta$. Applying differential privacy to neural networks helps ensure defenses against membership inference and model inversion attacks \cite{abadi2017privatetwoapproaches}. This can be achieved by either applying noise to gradients while training a single model \cite{abadi2016dpsgd} \cite{song2013stochastic} or by segregating data and adding noise in a collaborative learning setting \cite{papernot2018scalable} \cite{shokri2015privacy}. 

DP-SGD optimization was developed by \cite{abadi2016dpsgd} and involves adding Gaussian noise and clipping gradients of neural networks during training with stochastic gradient descent. It also keeps track of the privacy loss through a privacy accountant \cite{mcsherry2009privacyaccounting}, which prematurely terminates training when the total privacy cost of accessing training data exceeds a predetermined budget. Differential privacy is attained as clipping bounds the L2-norm of individual gradients, thus limiting the influence of each example on the learning updates.  We outline the DP-SGD algorithm briefly below in Algorithm \ref{algo_dpgradopt}.

\begin{algorithm}
\SetKwData{Left}{left}\SetKwData{This}{this}\SetKwData{Up}{up}
\SetKwFunction{Union}{Union}\SetKwFunction{FindCompress}{FindCompress}
\SetKwInOut{Input}{input}\SetKwInOut{Output}{output}
\Input{ Examples $\{x_1, . . . , x_N \}$, loss function $\mathcal{L}(\theta) = \frac{1}{N} \Sigma_i \mathcal{L} (\theta, x_i)$, Parameters: learning rate $\eta_t$, noise scale $\sigma$, group size $L$, gradient norm bound $C$.}
\Output{$\theta_T$ and calculate privacy cost $(\epsilon, \delta)$ using a privacy accountant method}
\textbf{Initialize} $\theta_0$ randomly\;
\For{$t \in [T]$}{
Take a random sample $L_t$ with sampling probability $L/N$\;
\textbf{Compute gradient}\;
For each $i \in L_t$, compute $\textbf{g}_t(x_i) \leftarrow \nabla_{\theta_t} \mathcal{L}(\theta_t, x_i)$\;
\textbf{Clip gradient}\;
$\overline{\textbf{g}}_t(x_i) \leftarrow \textbf{g}_t (x_i) / \text{max}(1, \frac{\lVert \textbf{g}_t (x_i) \rVert_2}{C}$\;
\textbf{Add noise}\;
$\Tilde{\textbf{g}}_t \leftarrow \frac{1}{L} (\Sigma_i (\overline{\textbf{g}}_t(x_i) + \mathcal{N}(0, \sigma^2 C^2 \mathcal{I})))$\;
\textbf{Descent}\;
$\theta_{t+1} \leftarrow \theta_t - \eta_t \Tilde{\textbf{g}}_t$\;
}
\caption{Differentially private SGD}\label{algo_dpgradopt}
\end{algorithm}\DecMargin{1em}

We use a modified DP-SGD algorithm \cite{abadi2016dpsgd} to fine-tune the entire network, using the following techniques introduced by \cite{zhang2018privatereleasing}; warm-starting, where a public dataset is used to initialize the weights of the model, and weights clustering, where the same dataset is used to estimate the gradient l2 norms of each parameter before using a hierarchical clustering algorithm to group parameters with similar clipping bounds (in Algorithm \ref{algo_weightcluster}).  The public dataset we use is a smaller retinal scan dataset from the STARE project \cite{hoover2000stare}.  We train the network privacy settings of $\epsilon=8.0$ and $\delta=1\times10^{-5}$ giving us a $\sigma$ value of $0.399$.  Finally, we retrain the layers to be encrypted as before, leaving us with our final model (DP-DFE-RN-50).

\begin{algorithm}
\SetKwData{Left}{left}\SetKwData{This}{this}\SetKwData{Up}{up}
\SetKwFunction{Union}{Union}\SetKwFunction{FindCompress}{FindCompress}
\SetKwInOut{Input}{input}\SetKwInOut{Output}{output}
\Input{k - target number of groups; $\{c(g_i)\}_i$ - parameter-specific gradient clipping bounds}
\Output{$\mathcal{G}$ - grouping of parameters}
$\mathcal{G} \leftarrow \{(g_i: c(g_i))\}_i$\;
\While{$|\mathcal{G}|$ $\geq$ k}{
$G, G' \leftarrow \text{arg min}_{G, G' \in \mathcal{G}} \text{max}(\frac{c(G)}{c(G')}, \frac{c(G')}{c(G)})$\;
merge $G$ and $G'$ w. clipping bound $\sqrt{c(G)^2 + c(G')^2}$
}
\textbf{return} $\mathcal{G}$\;
\caption{Weight Clustering}\label{algo_weightcluster}
\end{algorithm}\DecMargin{1em}

The accuracy and peformance metrics of our models are listed in Table \ref{tab:accuracy_results} and \ref{tab:performance_results}. We compare in terms of both a hardware-independent HOPS (homomorphic operations) and wall-clock runtime.

\subsection{Message Sizes}

The data transfer between the client and server consists of the values for the encrypted input and the values for the encrypted output prediction. Note that both the image and the prediction are encrypted under multiple keys held by the client, each corresponding to a distinct plaintext moduli $t^{(j)}$, which leads the message size to be proportional to the number of moduli used for  evaluation. We used fifteen plaintext moduli in our experiments.

As a result, the message size for the encrypted input in our DFE-ResNet-152 method is $15 \times 2 \times 2048 \times 7 \times 7 \times 8192 \times 4 \times 8$ bytes (789.2 GB), corresponding to the encryption of the $2048 \times 7 \times 7$ activation volume. The message size for the encrypted input in our CNN-8 baseline method is $15 \times 2 \times 224 \times 224 \times 3 \times 8192 \times 4 \times 8$ bytes (1183.8 GB), corresponding to the encryption of the $224 \times 224 \times 3$ input image. The message size of the encrypted output is identical in each case: $15 \times 2 \times 1 \times 8192 \times 4 \times 8$ bytes (7.9 MB).

Since the input is transformed to a representation with a smaller dimensionality in the transfer learning method, the cost of data transfer is reduced by 1.5x. While the message size is significant in both cases, we note that ciphertext batching techniques can amortize the cost of encrypted inference when a user wishes to request predictions on multiple images. In the case of diabetic retinopathy detection, this could correspond to predictions for both the left and right eye, or predictions for images of multiple patients of a healthcare provider.

\subsection{Experimental Correctness}
We validate on 100 images of our CIFAR-10 and Retina experiments to ascertain the correctness of our decrypted outputs.  Once more, we find around 0.05\% error due to the floating/fixed point conversion.

\subsection{Overall Comparison}
In a direct comparison between the baseline CNN-8 model and our transfer learning DFE-RN-50 model, we observe an across-the-board improvement.  DFE-RN-50 has higher accuracy, significantly reduces both the count of HOPs and measured runtime, and produces smaller message sizes than our baseline model.  We demonstrate the effectiveness of the sparsity-based optimization techniques to reduce computation time (7.8x speedup). We also show how other privacy concepts like differential privacy can be used to further improve the performance of our feature extraction architecture as we can see with the improved accuracy of DP-DFE-RN-50. To the best of our knowledge, this is the first implementation of homomorphic encryption and neural networks on a real-world medical imaging dataset.  

\section{Discussion}
Encrypted inference is not a panacea for private machine learning.  It has some obvious constraints the paper touches on in several sections, including the computational cost and network depth limitations.  Additionally, it does not cover the problem of private training and of defending against machine learning attacks.  The encrypted inference paradigm is still vulnerable to black box attacks, as it still returns encrypted outputs that are otherwise unaffected. For example, membership inference \cite{Fredrikson:2015:MIA:2810103.2813677} and model stealing \cite{tramer2016stealingapis} attacks can be performed with only access to the outputs of the model.  



\section{Conclusion}
Personal privacy is increasingly under threat in the modern digital age, and machine learning models continue to fuel the appetite for more invidual data and information. 
Homomorphic encryption holds great promise due to the security guarantees it can provide against both eavesdroppers and service hosts.
Unlocking its potential will require reducing the high overhead of arithmetic operations prevalent in neural networks.

In this work, we introduced and evaluated techniques for accelerating CryptoNets \cite{gilad2016cryptonets}.
The fundamental approach to our method is to leverage sparsity by using (i) efficient polynomial approximations for the activation functions and (ii) pruning and quantization that is tailored to the encryption scheme for significant performance gains. 
We show that our method, \textit{Faster CryptoNets}, is faster than CryptoNets without much loss of test set accuracy.
We also demonstrate how our technique can be deployed in a privately trained feature extraction setting, possibly inspiring future avenues of work where different privacy concepts can be combined to deliver an end-to-end privacy safe training and inference pipeline.  To the best of our knowledge, this is the first implementation of homomorphic encryption on a real-world medical imaging dataset.


Recent developments can produce even greater improvements.
Structured sparsity \cite{wen2016learning}, filter-level pruning methods \cite{luo2017thinet}, and efficient batching scheme and hardware acceleration techniques \cite{Migliore:2017:HAH:3145508.3126558} could further accelerate evaluation of deeper networks. 
In particular, more optimal encoding schemes could help reduce the message sizes of the encrypted data and provide more efficient parameters for the encryption scheme.
We hope this work will inspire future lines of research in efficient and privacy-safe machine learning.

\clearpage
\newpage

\bibliographystyle{IEEEtranS}

\end{document}